\documentclass[aps,twocolumn,nofootinbib,preprintnumbers,superscriptaddress,eqsecnum]{revtex4-1}

\usepackage{color}
\usepackage[normalem]{ulem}
\usepackage{amsmath}
\usepackage{enumerate}
\usepackage{amsfonts}
\usepackage{epsfig}
\usepackage[
      colorlinks=true,
      linkcolor=blue,
      urlcolor=blue,
      filecolor=black,
      citecolor=red,
      pdfstartview=FitV,
      pdftitle={},
        pdfauthor={},
        pdfsubject={},
        pdfkeywords={},
        pdfpagemode=None,
        bookmarksopen=true
      ]{hyperref}

\newcommand{\be}{\begin{equation}}
\newcommand{\ee}{\end{equation}}
\newcommand{\bea}{\begin{eqnarray}}
\newcommand{\eea}{\end{eqnarray}}

\newcommand{\bln}{\begin{align}}
\newcommand{\eln}{\end{align}}
\newcommand{\bst}{\begin{split}}
\newcommand{\est}{\end{split}}
\newcommand{\bi}{\begin{itemize}}
\newcommand{\ei}{\end{itemize}}
\newcommand{\ben}{\begin{enumerate}}
\newcommand{\een}{\end{enumerate}}

\def\le{\left}
\def\ri{\right}
\def\ha{{1\over 2}}
\def\lam{{\lambda}}
\def\Lam{{\Lambda}}
\def\Sig{{\Sigma}}
\def\al{{\alpha}}

\def\det{{\rm det}}

\def\th{{\theta}}

\def \th{{\theta}}

\def \lam {\lambda}
\def \om {\omega}

\def\sig{{\sigma}}
\def\ep{{\epsilon}}

\newcommand{\p}{\partial}

\newcommand\ga{{\ensuremath{{\gamma}}}}
\newcommand\Ga{{\ensuremath{{\Gamma}}}}

\def\lam{{\lambda}}

\def\eeq{\end{equation}}

\newcommand\sD{{\ensuremath{{\mathcal D}}}}

\newcommand\sL{{\ensuremath{{\mathcal L}}}}
\newcommand\sM{{\ensuremath{{\mathcal M}}}}
\newcommand\sN{{\ensuremath{{\mathcal N}}}}
\newcommand\sO{{\ensuremath{{\mathcal O}}}}

\newcommand\bxm{{\overline{x}_m}}
\newcommand\topj{{\widetilde{j}}}
\newcommand\tA{{\widetilde{A}}}
\newcommand\tphi{{\tilde{\phi}}}
\newcommand\tth{{\tilde{\th}}}
\begin{document}

\title {Monopole correlations in holographically flavored liquids}

\preprint{NSF-KITP-14-126}

\author{Nabil Iqbal}
\email{n.iqbal@uva.nl}
\affiliation{Kavli Institute for Theoretical Physics, University of California,
Santa Barbara, CA 93106, USA}
\affiliation{Institute for Theoretical Physics, University of Amsterdam, Science Park 904, Postbus 94485, 1090 GL Amsterdam, The Netherlands}


\begin{abstract}
Many-body systems with a conserved $U(1)$ current in $(2+1)$ dimensions may be probed by weakly gauging this current and studying correlation functions of magnetic monopole operators in the resulting dynamical gauge theory. We study such monopole correlations in holographic liquids with fundamental flavor, where the monopole operator is dual to a magnetically charged particle in the bulk. In charge-gapped phases the monopole operator is expected to condense. We show that this condensation is holographically dual to the capping off of the bulk flavor brane and compute the monopole condensate. We argue that from the lower-dimensional point of view this may be understood as a simple example of confinement of a gauge field in the bulk. In a compressible finite-density phase we present a novel calculation of the monopole correlation in space and time: the correlation is power law in time but is Gaussian in space due to interaction with the background charge density. 
\end{abstract}

\maketitle

\tableofcontents

\section{Introduction}\label{intro}

Consider a many-body system in $(2+1)$ dimensions with a global $U(1)$ current $j^{\mu}$. Different classes of infrared behavior for this current define different phases phases of matter. One useful way to probe this dynamics is to couple $j^{\mu}$ to an external gauge field $a_{\mu}(x)$,
\be
S_{QFT} \to S_{QFT} - \int d^3x \;a_{\mu}j^{\mu} \ , 
\ee
and study what happens as $a_{\mu}$ is wiggled slightly away from zero, allowing one to extract (for example) the correlation functions of the current. In this way one can understand very familiar observables such as the conductivity. 

In this paper we will consider a different sort of probe. Let us imagine weakly gauging the current $j^{\mu}$ by promoting $a_{\mu}$ to a {\it dynamical} gauge field. Now there is a new observable we can consider \cite{Borokhov:2002ib}: in the new functional integral over $a$, let us demand that its field strength $f$ have a {\it magnetic monopole} singularity at a Euclidean spacetime point $x_m$:
\be
df = q_m \delta^{(3)}(x - x_m) \label{sketchydef}
\ee
where $f = da$. This operation deforms the theory at the point $x_m$ and may be understood as inserting a local operator $\sM(x_m)$ at that point:
\be
\langle \sM(x_m) \rangle \equiv Z^{-1} \int_{x_m} [\sD a \sD \phi] \exp\le(-S_{QFT}[\phi] + \int d^3x j^{\mu} a_{\mu}\ri) \, \label{locins}
\ee
where $Z$ is the undeformed partition function and the subscript $x_m$ indicates that the functional integral over $a_{\mu}$ obeys the modified boundary condition \eqref{sketchydef}. 
$\sM(x)$ is called a {\it monopole operator} and is an example of a more general class of disorder operators which are defined not as polynomials of fundamental fields appearing in the Lagrangian but rather by the modification of boundary conditions obeyed by such fields in the path integral, as in \eqref{sketchydef}. 

Having defined $\sM$ we can study its two-point function $\langle \sM(x) \sM^{\dagger}(0) \rangle$: this {\it monopole correlation function} measures how the system responds to an injection of magnetic flux, and will be a nontrivial function of the separation $x$ between the monopole and anti-monopole. Importantly, the magnetic monopole charge $q_m$ satisfies a Dirac quantization condition and cannot be made arbitrarily small. Thus the configuration above necessarily represents non-perturbative information that is {\it not} contained in simple correlation functions of the current $j^{\mu}$, as will be increasingly clear as we proceed. 

There are various reasons to study this object. Many interesting phases in conventional condensed matter physics result in the presence of an emergent dynamical $U(1)$ gauge field that one may identify with $a$ (e.g. the $U(1)$ spin liquids; for reviews see \cite{RevModPhys.78.17,Wen:2004ym} and references therein). As originally shown by Polyakov, the presence of magnetic monopoles can dramatically change the infrared physics of a dynamical gauge theory: if the local operator $\sM(x)$ that one identifies with the monopole insertion is {\it relevant}, then its presence will drive the system to a new IR phase in which the $U(1)$ gauge field is confined \cite{Polyakov:1975rs, Polyakov:1976fu}. The presence of gapless charged matter directly affects this monopole dimension: thus it is of intrinsic interest to study monopole correlations in different phases of matter. As we review below, the non-perturbative nature of this object makes it difficult to compute using conventional field-theoretical techniques in all but the most symmetric of settings. 

Independent of such considerations, we will argue that the monopole correlation is an interesting probe of phases of matter in its own right, being sensitive to the IR structure of the charge sector in a novel manner. As a probe of gapless charged matter, it appears to be ideally suited to a holographic description. The boundary monopole operator associated to a current $j$ is dual to a bulk particle that is {\it magnetically} charged under the bulk gauge field $A$ that is dual to $j$ \cite{Witten:2003ya,Sachdev:2012tj}. Its correlation function can be given a simple geometric interpretation in the bulk and is easily computable. Such bulk monopoles have previously been argued to be important for a fine-grained understanding of holographic matter in \cite{Faulkner:2012gt,Sachdev:2012tj}.  In this paper we will study them in a particular top-down holographic model where the $U(1)$ current in question will be the baryon number associated with fundamental flavor degrees of freedom. While some of our considerations will apply to any holographic model, we will see that the extra control associated with the top-down construction will be useful at certain points. 

We now present a brief outline of this note. In Section \ref{sec:ftmoncorr} we review field-theoretical results for monopole correlations in various settings. In Section \ref{sec:holflav} we turn to holography and introduce the particular brane embedding that we will study, as well as the holographic realization of the monopole operator as a wrapped D-brane. In Section \ref{sec:gapped} we turn to a holographic phase with a charge gap. Here one expects the monopole operator to {\it condense}: we demonstrate this condensation geometrically from the bulk. We also argue that a bulk charge gap can be understood as confinement of the {\it bulk} U(1) gauge field, as anticipated in \cite{Faulkner:2012gt,Sachdev:2012tj}. In Section \ref{sec:finiterho} we compute the monopole correlation function on a compressible phase with a finite density of the $U(1)$ charge $\rho$. We find a Gaussian suppression of the correlation function due to the interaction of the monopole with the background electric field. We conclude in Section \ref{sec:conc} with a brief discussion and some directions for future research. 

Previous discussion of magnetic monopoles in AdS includes a realization of monopoles as solitons \cite{Bolognesi:2010nb,Sutcliffe:2011sr}. The implementation of the monopole as a wrapped D-brane has also been discussed in a slightly different context in the recent work \cite{Filev:2014mwa}.

\section{Monopole correlations in field theory} \label{sec:ftmoncorr}
In this section we discuss in more detail the construction of the monopole correlation function and review some expectations for this object from field theory. As described above, given a theory with a global $U(1)$ current $j^{\mu}$, we may couple the current to a dynamical gauge field $a_{\mu}$ and then define the monopole correlation function to be 
\begin{align}
& \langle \sM(x_m) \sM^{\dagger}(\bxm) \rangle \equiv \nonumber \\ & Z^{-1} \int_{(x_m,\bxm)} [\sD a \sD \phi] \exp\le(-S_{QFT}[\phi] + \int d^3x j^{\mu} a_{\mu}\ri) \, \label{moncorr}
\end{align}
where $\phi$ denotes the underlying fields of the QFT and the subscript $x_m$, $\bxm$ indicates that all $a_{\mu}$ in the functional integral should satisfy the following monopole boundary condition, corresponding to placing a monopole at $x_m$ and an anti-monopole at $\bxm$,
\be
df = q_m \le(\delta^{(3)}(x - x_m) - \delta^{(3)}(x - \bxm)\ri) \ .\label{twomonop}
\ee
This monopole correlation function will be the main object of study in the remainder of this note. 

\begin{figure}[h]
\begin{center}
\includegraphics[scale=0.35]{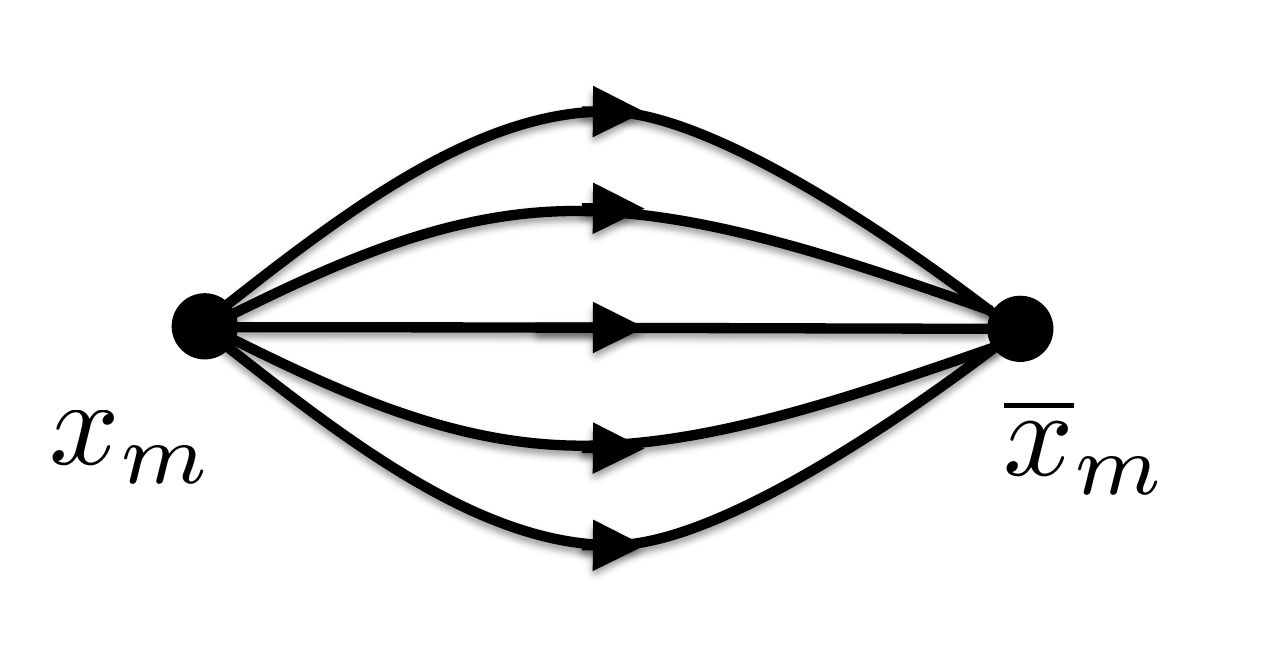}
\end{center}
\vskip -0.5cm
\caption{Schematic of field lines around monopole antimonopole pair.}
\label{fig:monantimon}
\end{figure}

One may be uneasy about the definition \eqref{moncorr} without a bare kinetic term for $a_{\mu}$. As it turns out, as we flow to the IR the coupling of $a_{\mu}$ to the current will always introduce an effective kinetic term and there is no need to specify one in the UV. As our motivation is to study an interesting probe of the original theory (with no dynamical gauge field), one might also worry that the fluctuations of $a_{\mu}$ could drive us to a new IR phase that is very different from the original one. While this is a concern in principle, we will ignore it in this note, assuming that the large number of charged degrees of freedom $N \gg 1$ that are present in all our models will effectively suppress fluctuations of $a_{\mu}$\footnote{As discussed in \cite{Pufu:2013eda}, in conformal theories it is also possible to define a monopole operator for a strictly global $U(1)$ symmetry, i.e. without making $a_{\mu}$ dynamical, essentially by fixing a suitable monopole profile for $a_{\mu}$. While these objects are very similar to those studied in this paper, they differ at subleading order in $1/N$ (when fluctuations of $a_{\mu}$ begin to play a role) and appear difficult to precisely formulate in non-conformal theories.}. Essentially we are simply asking how the field theory responds to a gauge field source of the form shown in Figure \ref{fig:monantimon}, where the path integral over $a_{\mu}$ allows it to relax to a configuration that minimizes the action subject to the monopole boundary conditions. 

There is another way to understand the monopole correlator. Any $U(1)$ gauge theory in $(2+1)$ dimensions (such as the one we have just constructed) has a topological global $U(1)$ current $\topj^{\mu}$:
\be
\topj^{\mu} \equiv \frac{i}{4\pi} \ep^{\mu\nu\rho} f_{\nu\rho}, \label{topjdef}
\ee
whose conservation follows from the Bianchi identity. However, we see directly from the definition of the monopole operator $\sM$ in \eqref{locins} that this current is not {\it quite} conserved in its presence:
\be
\langle \p_{\mu} \topj^{\mu}(x) \sM(y) \cdots \rangle = \frac{i q_m}{2\pi} \delta^{(3)}(x-y)\langle \sM(y) \cdots \rangle \label{ward}
\ee
This expression is of precisely the correct form to be a Ward identity for the topological current $\topj^{\mu}$: thus we conclude that the monopole operator $\sM(x)$ has a charge $\frac{q_m}{2\pi}$ under the topological current. Questions about the long-distance behavior of the monopole correlator can be understood in terms of the realization of the global symmetry generated by $\topj^{\mu}$. We now briefly review expectations for the monopole correlator from field theory. 
\subsection{Conformal matter} \label{sec:confft}
We begin with the most symmetric setting, a conformal field theory. Here we expect the correlator to be a power law,
\be
\langle \sM(x) \sM^{\dagger}(0) \rangle \sim \frac{1}{|x|^{2 \Delta}} \ . \label{powerL}
\ee
The correlation is completely determined by the dimension $\Delta$. The most efficient way to compute this dimension is to perform a conformal mapping to $S^2 \times \mathbb{R}$ \cite{Borokhov:2002ib}. A monopole insertion at the origin with magnetic charge $q_m$ corresponds to having $q_m$ units of magnetic flux on the $S^2$. The Casimir energy of the resulting field theory state is related to the dimension of the monopole operator via the usual state-operator correpondence. In the limit of a large number of charged degrees of freedom, one can neglect gauge field fluctuations and simply compute the energy of the charged degrees of freedom on a monopole background. This computation has been performed explicitly in a handful of conformal field theories. For example for a unit charged monopole in a theory of $N_f$ free fermions one finds \cite{Borokhov:2002ib}
\be
\Delta_f = N_f \cdot 0.265 \ \label{freeferm}, 
\ee
and for free and critical bosons one finds \cite{Pufu:2013eda,Murthy:1989ps,PhysRevB.78.214418}:
\begin{align}
\Delta_{free} & = N_b \cdot 0.097 \label{freebose}\\ 
\Delta_{critical} & = N_b \cdot 0.125 \ . 
\end{align}
Further work, including the computation of $\sO(1)$ corrections and the generalization to supersymmetric and non-Abelian theories, can be found in \cite{Pufu:2013vpa,Borokhov:2002cg,Dyer:2013fja}. 
\subsection{Gapped charge carriers} \label{sec:gapcharge}
We turn now to the case of gapped charges. We can no longer use the conformal mapping, and in fact it is now very difficult to obtain an explicit expression for the monopole correlator $\langle \sM(x) \sM^{\dagger}(0) \rangle$ as a function of the separation, even for the free theory obtained in a large $N$ limit. However we may qualitatively understand the extreme IR limit: in the infrared, all the charges are gapped, and we can integrate them out and obtain an effective action for $a$. Assuming parity, the most relevant contribution will be the Maxwell term:
\be
\Ga[a] = \frac{1}{4 M} \int d^3x\;(da)^2 \label{maxwell}
\ee
where the effective IR gauge coupling $\sqrt{M}$ is set by the mass of the charged degrees of freedom. 

This actually describes a phase where the topological symmetry generated by $\topj$ in \eqref{topjdef} is {\it spontaneously broken}. To understand this, we first add a source term $B_{\mu} \topj^{\mu}$ to the Maxwell action. Now free Maxwell theory in $(2+1)$ dimensions is equivalent to the theory of a free compact scalar $\phi$; performing the usual duality (e.g. as in Appendix B of \cite{polchinski1998string}) with the source term added we find the action
\be
\Ga[\phi;B] = \frac{M}{2} \int d^3x \le(\nabla\phi - \frac{B}{2\pi}\ri)^2 \label{phiac}
\ee
where the relation between $f$ and $\phi$ is $f^{\mu\nu} = -i M \ep^{\rho\mu\nu}\le(\p_{\rho}\phi - \frac{B_{\rho}}{2\pi}\ri)$. The action \eqref{phiac} describes a theory where the current $\topj$ that couples to the external source $B$ is spontaneously broken, with $\phi$ the corresponding Goldstone mode. This duality is possible only when there is no charged matter coupled to $a_{\mu}$ and thus should apply only at scales much longer than $M^{-1}$. We conclude that a charge gap implies that the topological symmetry is broken in the infrared. 

What does this imply for the monopole correlator? We see from \eqref{ward} that that the monopole operator $\sM$ is charged under $\topj$ and so should act as an order parameter for the symmetry breaking. Thus in a phase with a charge gap we expect the correlator to saturate
\be
\lim_{x \to \infty} \langle \sM(x) \sM^{\dagger}(0) \rangle \sim \langle \sM \rangle^2 \neq 0 \, \label{monsat}
\ee
where the monopole condensate should be set by the charge gap $\langle \sM \rangle \sim M^{\Delta}$. In particular, if we deform a conformal theory by adding a mass term we expect the monopole correlator to interpolate between \eqref{powerL} at short distances and a constant \eqref{monsat} at long distances. 
\subsection{Superfluid} \label{sec:superfluid}
We turn now to a superfluid, when the symmetry generated by the {\it original} $U(1)$ current $j^{\mu}$ has been spontaneously broken by a condensate of bosons with charge $q_e$. For simplicity first consider a relativistic system with zero net charge, $\rho = 0$. Upon gauging by $a_{\mu}$ the system becomes a superconductor with effective action
\be
\Ga[\th,a] = \int d^3x\;\rho_s\le(\nabla\th - q_e a\ri)^2 \label{sfac}
\ee
where $\th$ is the Goldstone mode associated with the breaking of $j^{\mu}$. 

\begin{figure}[h]
\begin{center}
\includegraphics[scale=0.5]{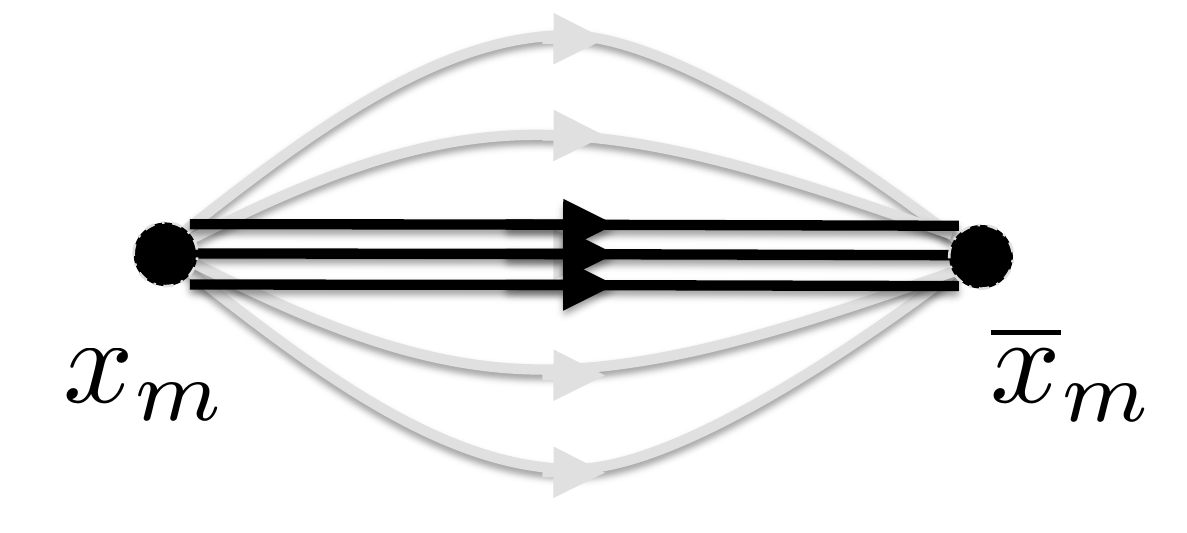}
\end{center}
\vskip -0.5cm
\caption{In a superfluid phase field lines are forced into an Abrikosov-Nielsen-Olesen vortex (solid black lines) rather than being able to spread out (gray lines). The vortex may be viewed as the worldline of a massive particle that is charged under $\topj$.}
\label{fig:superflux}
\end{figure}

Consider now inserting a monopole operator. The magnetic flux created by this monopole cannot propagate freely in the superconductor; the ordinary Meissner effect will force the flux lines into a single Abrikosov-Nielsen-Olesen vortex. Thus a monopole-antimonopole pair is always connected by a flux tube with finite tension $T$ as in Figure \ref{fig:superflux}, and so the correlator should decay (at least) exponentially in space:
\be
\langle \sM(x) \sM^{\dagger}(0) \rangle \sim \exp\le(- T|x|\ri) \ . \label{sfans}
\ee
Equivalently, the monopole operator $\sM$ creates a massive state -- the Abrikosov-Nielsen-Olesen vortex -- which carries charge $\frac{q_m}{2\pi}$ under $\topj$. This is the only excitation charged under $\topj$: as it is massive, the symmetry generated by $\topj$ is unbroken. 

If the net charge is nonzero we actually expect the correlation function to decay faster than exponential: we will return to this issue in the conclusion. 

\subsection{Fermi liquid} \label{sec:FL}
Consider now a Fermi liquid, which is a compressible phase that contains a finite density of charge $\rho$ {\it without} breaking the $U(1)$ symmetry. It also contains many low-lying degrees of freedom at the Fermi surface, far more than the single Goldstone mode studied above. Here there are no simple techniques to understand how these excitations affect the monopole correlation function: we have neither relativistic conformal symmetry nor a simple understanding of monopoles as creating gapped vortices as in the superfluid case. There is however a (generically uncontrolled) method that has been used in the literature \cite{PhysRevB.78.045110,PhysRevB.50.8078,PhysRevB.68.195110}, the one-loop or RPA approach, which we take a moment to outline here. We stress that this approach does not actually result in quantitatively correct results \cite{PhysRevB.70.214437}, though one might hope for qualitative agreement. Our main motivation is to contrast this with the holographic calculation that will follow. 

Consider taking the (unperturbed) Fermi liquid state and integrating out the low-lying modes near the Fermi surface to obtain an effective quadratic action for the gauge field $a_{i}$:
\be
\Ga[a] = \int \frac{d\om d^2k}{(2\pi)^3} a_i(\om,k) a_j(-\om,-k) K^{ij}(\om,k) \label{effA}
\ee
$K_{ij}$ is a kernel that is the current-current correlation function $\langle j_i(k) j_j(-k)\rangle$. As we have integrated out (many) gapless degrees of freedom the kernel is quite nonlocal in position space. Given this effective action, it might seem reasonable to simply vary it with respect to $a_i$, subject to the monopole boundary conditions \eqref{twomonop}. After this we can compute the action of the monopole-antimonopole pair on-shell, and the correlation function would be
\be
\langle \sM(x) \sM^{\dagger}(0) \rangle \sim \exp\le(-\Ga[x]\ri) \ . 
\ee
This procedure has been carried out in the references above for the Fermi liquid with different sorts of interactions (i.e. different bare kinetic terms for the gauge field). While there is some variation in the literature depending on the precise model used, one typically finds answers that fall off exponentially in space (i.e. superficially similar to \eqref{sfans}, but with a different physical origin).  

The serious problem with this approach is that it treats the problem perturbatively in the size of the monopole field, whereas the monopole is an intrinsically non-perturbative object \cite{PhysRevB.70.214437}, as the magnitude of its field $q_m$ satisfies a Dirac quantization condition. We cannot justify dropping the terms that are higher order in $a$ in \eqref{effA}. For example, applied to the free relativistic fermion this technique results in a power-law correlation as in \eqref{powerL}, but with a $\Delta \sim q_m^2$ as befitting a classical interaction energy. However this is wrong: the actual dependence of $\Delta$ on the {\it discrete} parameter $q_m$, computed using CFT techniques in \cite{Borokhov:2002ib,Pufu:2013vpa}, is not quadratic.  

Importantly, it has however been argued nonperturbatively that the monopole dimension is formally ``infinite'' with respect to the scaling symmetry that scales single-particle modes towards the Fermi surface \cite{PhysRevB.78.085129}. This means that the monopole operator is not relevant and suggests that the correlation function should fall off faster than a power-law, but we do not know of a controlled field-theoretical method to compute the actual scaling function. 

The nonperturbative nature of the monopole operator will be particularly clear in the holographic computations that follow, and we will revisit this issue in the conclusion. 

\section{Holographic flavor and monopole operators} \label{sec:holflav}
Having reviewed expectations from field theory, we now turn to holography. Given a $(2+1)$ dimensional field theory with a gravitational dual, we seek to understand the bulk object that is dual to the monopole operator $\sM(x)$. Soon we will specialize to a particular field theory, but first we make some general statements that should apply to any holographic model.  

The conserved current $j^{\mu}(x)$ is dual to a gauge field $A_M(r,x)$ in the (3+1) dimensional bulk. Now a local operator on the boundary is generally dual to a propagating field in the bulk. If this field has a large mass -- which will turn out to be the case for the monopole operator -- then the physics is well described by the individual quanta of the field, i.e. by particles. It is well understood that if such a bulk particle is {\it electrically} charged under $A_M$ then it is dual to an operator that carries charge under $j^{\mu}$. It should then seem very natural that a particle that is {\it magnetically} charged under $A_M$ -- i.e. a {\it bulk} magnetic monopole -- is dual to the monopole operator $\sM(x)$, which carries ``magnetic charge'' with respect to $j^{\mu}$ in the manner defined above. Some details related to this identification can be found in \cite{Sachdev:2012tj}, including the holographic computation of the three-point function between $j$ and the monopole operators.  

We present a quick way to understand this. Consider the definition \eqref{locins} of the monopole operator. In AdS/CFT, the external gauge field source $a_{\mu}$ is identified with the boundary value of the bulk gauge field $A_\mu(r \to \infty, x) = a_{\mu}(x)$. In the large-$N$ limit, the path integral over $a_{\mu}$ reduces to the tree-level demand that the bulk action be stationary with respect to variations of the boundary value of $A_{\mu}$, subject to the boundary condition \eqref{locins}. 

\begin{figure}[h]
\begin{center}
\includegraphics[scale=0.5]{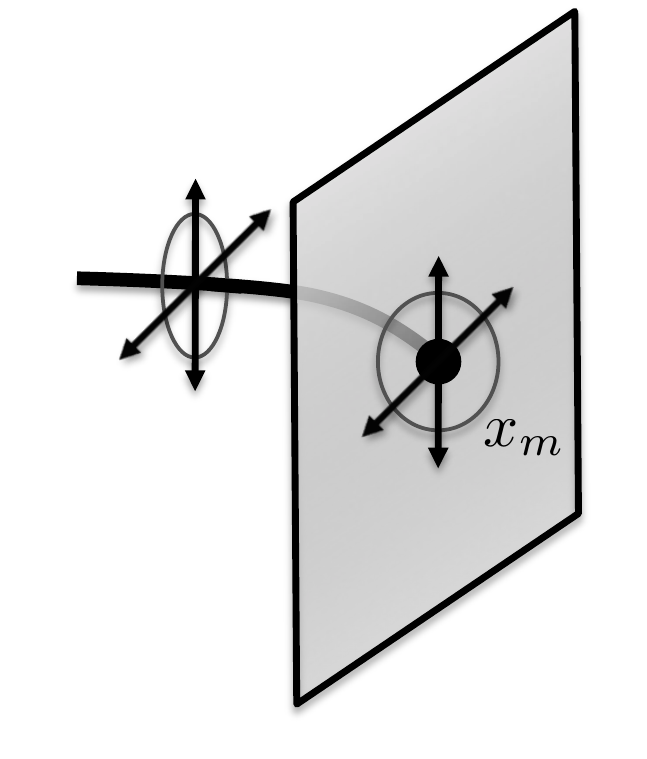}
\end{center}
\vskip -0.5cm
\caption{Intersection of bulk monopole worldline with boundary is insertion of field theory monopole operator at $x_m$. Note any $S^2$ surrounding the bulk worldline will register a nonzero magnetic flux.}
\label{fig:monopolebdy}
\end{figure}

Now imagine a small $S^2$ at the boundary surrounding the monopole insertion point $x_m$, as shown in Figure \ref{fig:monopolebdy} We have
\be
\int_{S^2} dA(r \to \infty) = \int_{S^2} da = q_m 
\ee
As $a$ is a dynamical gauge field, this boundary condition indicates the presence of a defect sourcing the gauge field at $x_m$ on the boundary. This source can only be supplied by the bulk magnetic monopole, whose worldline must now intersect the boundary at $x_m$. This is precisely the statement that the bulk monopole is the dual of the local operator defined by \eqref{locins}. Note that if we we now pull the $S^2$ off the boundary and move it into the bulk, the nonzero flux will persist whenever the $S^2$ surrounds a one-dimensional curve $C$ -- the worldline of the monopole. 

These considerations apply to any reasonably consistent example of holography, in particular to bottom-up models. However, we will see that some of the physics that we are interested in will require extra information (i.e. the completion provided by string theory) for a controlled description. 

\subsection{D3D5 intersection}
To that end we now specialize to a particular field theory, the well-studied D3D5 intersection. Here we will take $N_c$ D3 branes and a single D5 brane intersecting along $(2+1)$ dimensions. The field theory consists of $\sN = 4$ Super Yang-Mills with gauge group $SU(N_c)$ in $(3+1)$ dimensions from the D3 branes. The D5 brane contributes matter charged in the fundamental under this $SU(N_c)$ but living on a $(2+1)$ dimensional defect \cite{DeWolfe:2001pq,Erdmenger:2002ex,Karch:2002sh}\footnote{See (e.g.) Chapter 8 of \cite{CasalderreySolana:2011us} for an introductory review of fundamental flavor in AdS-CFT via probe branes.}. We will focus only on the dynamics that is localized on this defect. At zero coupling the field-theory action for the degrees of freedom on the defect is
\be
S_{defect} = \int d^3 x\le(|D_{\mu} q|^2 - i \overline{\Psi} \ga^{\mu} D_{\mu} \Psi\ri)
\ee
Here the $q$'s are complex scalars that transforms in the fundamental under a global symmetry $SU(2)_H$, while the Dirac fermions $\Psi$ transform in the fundamental of a different global symmetry $SU(2)_V$. Both the scalars and fermions are in the fundamental of the gauge group $SU(N_c)$. The $U(1)$ current that we will study is the baryon number current $U(1)_B$, which acts as a phase rotation on both $\Psi$ and $q$. This system has been very well-studied as an example of top-down holography in $(2+1)$ dimensions and we briefly review it here. 

At strong coupling the D3 branes coalesce into the usual $AdS_5 \times S^5$, with the standard relations for the AdS radius $R$ and the string coupling in terms of gauge theory quantities:
\be
g_s = \frac{g_{YM}^2}{4\pi} \qquad \frac{R}{l_s} = (g_{YM}^2 N_c)^\frac{1}{4} \ . 
\ee
The single D5 brane can be treated as a probe brane: neglecting the backreaction of the probe is dual to neglecting the effect of the $\sO(N_c)$ fundamental matter degrees of freedom on the $\sO(N_c^2)$ gluons. 

To determine the possible phases we minimize the DBI action of the D5 brane
\be
S_5 = -T_5 \int d^6 \sig \sqrt{-\det\le(\ga_{5} + 2\pi \al' F\ri)}, \label{DBI}
\ee
where $\ga_5$ is the induced metric on the D5 brane worldvolume, and $F = dA$ is the field strength of the bulk gauge field $A$ that is dual to the $U(1)_B$ current $j$. The tension of a Dp brane is
\be
T_p = \frac{1}{(2\pi)^p g_s l_s^{p+1}} \ . \label{Dptension}
\ee
We take the metric of a unit $S^5$ to be
\begin{align}
ds_{S^5}^2 = d\psi^2 & + \sin^2\psi\le(d\th^2 + \sin^2 \th d\phi^2\ri) \nonumber \\ 
& + \cos^2 \psi\le(d\tilde{\th}^2 + \sin^2 \tilde{\th} d\tilde{\phi}^2\ri) \ , \label{S5metric}
\end{align}
and we write the metric of $AdS_5$ as 
\be
ds_{AdS_5}^2 = \frac{R^2}{z^2}\le(-dt^2 + dx^2 + dy^2 + dx_{\perp}^2 + dz^2\ri),
\ee
There is a solution where the D5 brane is extended in the $\th$ and $\phi$ directions and so wraps the first $S^2 \subset S^5$, sitting at the point $\psi = \frac{\pi}{2}$. $SU(2)_H$ acts as $SO(3)$ rotations on $(\th, \phi)$, and  $SU(2)_V$ acts as $SO(3)$ rotations on $(\tth, \tphi)$. The D5 brane also sits at $x_\perp = 0$; the remaining four dimensions $(t,x,y,z)$ form an $AdS_4$ slice inside the bulk $AdS_5$:
\be
ds_{AdS_4}^2 = \frac{R^2}{z^2}\le(-dt^2 + dx^2 + dy^2 + dz^2\ri) \ . \label{AdS4met}
\ee 
This $AdS_4$ indicates that the dual $(2+1)$ dimensional defect theory is conformally invariant. 

There are other possibilities for the IR dynamics: indeed we may realize many of the possibilities discussed in Section \ref{sec:ftmoncorr} above, but for the remainder of this section we will study the conformal phase.  

\subsection{Monopole operator} \label{sec:monop}
We now need to identify the bulk object that is dual to the monopole operator $\sM(x)$. Following the discussion above, this should be a particle (i.e. a 1-dimensional object in spacetime) in $AdS_4$ and is magnetically charged under the worldvolume gauge field $A$.

\begin{figure}[h]
\begin{center}
\includegraphics[scale=0.5]{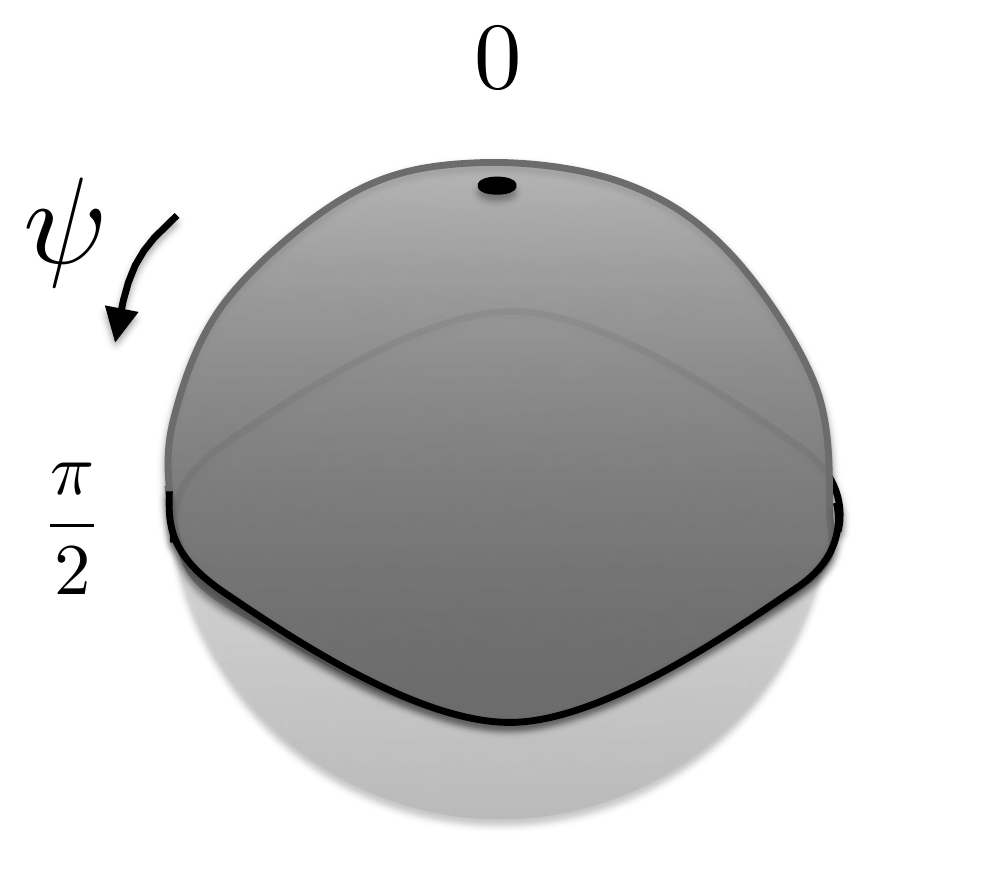}
\end{center}
\vskip -0.5cm
\caption{D3 brane ends on D5 brane worldvolume (at $\psi = \frac{\pi}{2}$), filling in half-$S^3$ and extending from $\psi = 0$ to $\psi = \frac{\pi}{2}$.} 
\label{fig:wrappedD3}
\end{figure}

The correct object is a wrapped D3 brane that {\it ends} on the D5 brane. The boundary of the D3 brane is a (2+1) dimensional manifold: as the D3 ends on the D5, this boundary must lie on the D5 worldvolume. Take this boundary to wrap the compact $S^2$: as the $S^2$ shrinks to zero size inside the $S^5$, the D3 brane can now {\it fill in} a half $S^3$, as shown in Figure \ref{fig:wrappedD3}. In the coordinates of \eqref{S5metric} the D3 brane is extended in the $\psi$ direction from $\psi = 0$ to $\psi = \frac{\pi}{2}$ where it ends on the D5 brane. The remaining one dimension of the D3 brane becomes a worldline $C$ on the $AdS_4$. This D3 brane configuration and its identification as a magnetic monopole has recently been studied in \cite{Filev:2014mwa}.

Branes ending on branes appear as magnetic sources to worldvolume fields when the difference in dimension is $2$ \cite{Strominger:1995ac}. In Appendix \ref{app:effac} we work out the couplings between this worldline and the worldvolume fields on the D5 and show that indeed there is a coupling of the form 
\be
S_{D3, \widetilde{A}} = 2\pi \int_C \widetilde{A} \label{magcoup}
\ee
where $\widetilde{A}$ is the magnetic dual of the worldvolume gauge field $A$. It satisfies
\begin{align}
& (d \widetilde{A})_{MN} \nonumber \\
& = \sN \sqrt{-\det_4\le(\ga + 2\pi \al' F\ri)}[(\ga + 2 \pi \al' F)^{-1}]^{PQ}\ep_{PQMN} \label{magdual}
\end{align}
with the normalization $\sN = -4 \pi^2 R^2 \al' T_5$. 

This may look somewhat unfamiliar, but is actually the generalization of the usual idea of electric-magnetic duality to the nonlinear kinetic term of the DBI action \eqref{DBI}. Importantly, the right hand side of the above expression is the object that obeys Gauss's law, and the constraint that $d(d\widetilde{A}) = 0$ is equivalent to the dynamical equation of motion for $A$, as we expect for electric-magnetic duality. To first order in $F$ this is the familiar Maxwell expression $d\widetilde{A} \sim \star_4 F$. Note that $\widetilde{A}$ is dual to the boundary topological current $\topj$ \cite{Witten:2003ya}. The charge quantum in \eqref{magcoup} corresponds to a magnetic charge that saturates the Dirac quantization condition, where the unit electric charge is taken to be the endpoint of a fundamental string.  

To compute the two-point function of the monopole operator we should demand that this D3 brane intersect the $AdS_4$ boundary at two points separated by a distance $\Delta x$. The on-shell action of the brane will determine the correlation function:
\be
\langle \sM(\Delta x) \sM^{\dagger}(0) \rangle \sim \exp\le(-S_{D3}[\Delta x]\ri) \ . \label{corrfunc}
\ee
This action has two parts: a geometric portion given by the DBI action of the D3 brane and a coupling to the worldvolume gauge field given by \eqref{magcoup}. We now compute the effective mass of the wrapped D3 brane from the $AdS_4$ point of view. The DBI action for the D3 brane is
\be
S_{D3} = -T_3 \int_{D3} d^4\sig \sqrt{-\ga_3} = - m\int_C ds
\ee
where to obtain the effective 4d mass we integrate the tension \eqref{Dptension} over the half $S^3$ to find
\be
m = \frac{R^3}{g_s l_s^4} \frac{4\pi}{(2\pi)^3}\int_0^{\frac{\pi}{2}} d\psi \sin^2 \psi = \frac{N_c}{2R} \ .  \label{D3mass}
\ee
Using the usual (large-mass) AdS/CFT relation $\Delta = mR$ we conclude that in this theory the dimension of the unit-charge monopole operator $\sM(x)$ at strong coupling is
\be
\Delta = \frac{N_c}{2} \ . \label{Delta}
\ee
As expected, the dimension essentially counts the number of charged degrees of freedom\footnote{This is similar to the monopole dimension computed in the supersymmetric model of \cite{Borokhov:2002cg}: in our gauging of $U(1)_B$ we have not attempted to preserve supersymmetry, but it seems likely that there exists a supersymmetric cousin to our monopole (perhaps one with other fields turned on) whose dimension could be derived from a field-theoretical analysis.}. We can compare this to the dimension at weak coupling: using the results for free bosons and fermions in \eqref{freeferm} and \eqref{freebose} we find the qualitatively similar $\Delta_{free} \approx N_c \cdot 0.724$. By conformal invariance, if we compute the two-point function using \eqref{corrfunc} we will find that the correlator is a power law as in \eqref{powerL} with $\Delta$ given by \eqref{Delta}. 

The monopole operator is characterized by other quantum numbers than simply its dimension. In particular, it transforms in a spin $\frac{N_c}{2}$ representation under the global R-symmetry $SU(2)_V$. This can be understood field-theoretically from the presence of fermion zero modes bound to the monopole, and holographically from the quantization of the movement of the D3 brane in the $(\tilde{\th}, \tilde{\phi})$ directions. The construction of this representation is interesting but unrelated to our main narrative, and so we relegate it to Appendix \ref{app:global}. 

Finally, we note that at the attachment points the D3 brane pulls on its parent D5 brane, sourcing the field $\psi$. Taking this to its logical conclusion, there is an alternative BIon-like \cite{Callan:1997kz} description of the attached D3 brane in terms of D5 brane fields alone: rather than attach an explicit D3 brane, we can instead construct a D5 brane profile where the internal $S^2$ (i.e. whose size is parametrized by $\psi)$ shrinks but the $S^2$ surrounding the curve $C$ in $AdS_4$ remains at a finite size, allowing us to put magnetic flux on it without requiring an explicit source. This is somewhat like a Minkowski embedding for the D5 brane, and this configuration has recently been studied in detail in \cite{Filev:2014mwa}. It mimics the geometry of the attached D3 brane. We have checked that in the limit of a small amount of flux, the net energy of the resulting configuration is precisely the same as that computed above from the D3 brane alone, as is often the case for such BIons. At our level of description there appears to be little difference between considering the excitation in question to be a BIon or an attached D3 brane.

\section{Gapped charges} \label{sec:gapped}
We now discuss a phase where we deform the CFT away from criticality. The operator dual to the slipping mode $\psi(z)$ -- which, as shown in \eqref{S5metric}, controls the size of the $S^2$ that the D5 brane wraps -- has dimension $\Delta_{\psi} = 2$ and so is a mass term in the CFT. If we turn on this source in the UV then $\psi(z)$ develops a radial profile in the bulk. The effective action for $\psi$ can be worked out from \eqref{DBI} to be
\be
S_{D5} = -4\pi T_5 R^6\int dz \le(\frac{\sin^2\psi(z)}{z^4}\sqrt{1+ z^2\le(\frac{\p \psi(z)}{\p z}\ri)^2}\ri),
\ee
whose equations of motion admit the following exact solution
\be
\psi(z) = \cos^{-1}\le(\frac{z}{z_m}\ri),
\ee
with $z_m$ a constant of integration \cite{Karch:2002sh}. Note that we have $\psi(z_m) = 0$: thus at this value of the radial coordinate, the $S^2$ has closed up entirely and the brane has smoothly capped off in a Minkowski embedding, indicating the presence of a mass gap for all flavored degrees of freedom, and in particular for any excitation that is charged under the global $U(1)_B$ symmetry. Further, expanding $\psi(z)$ near the boundary we have
\be
\psi(z \to 0) \sim \frac{\pi}{2} - \frac{z}{z_m} + \sO(z^3) \ .
\ee
The term linear in $z$ is the source, and so we see the bare boundary mass $M = z_m^{-1}$. The absence of a quadratic term indicates that there is no condensate of the operator dual to $\psi$. 
\begin{figure}[h]
\begin{center}
\includegraphics[scale=0.55]{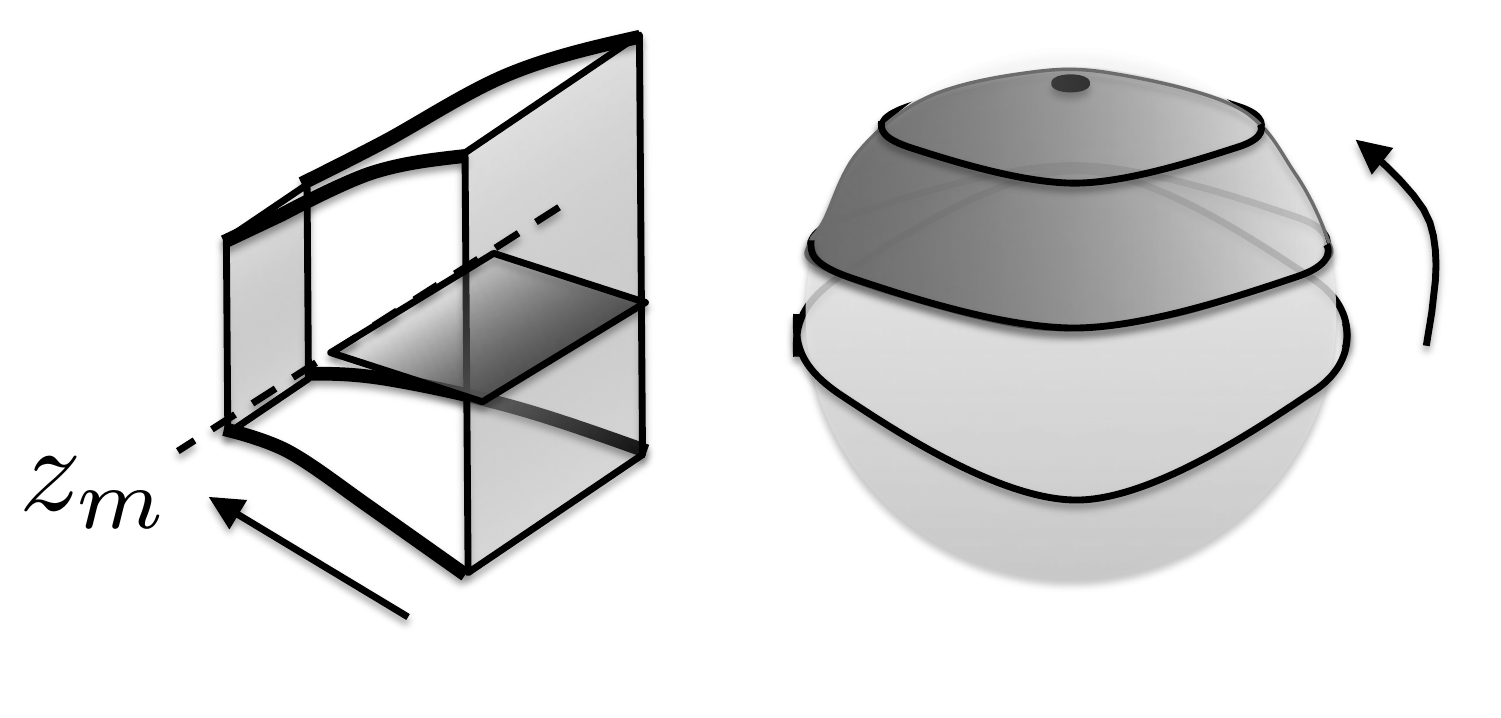}
\end{center}
\vskip -0.5cm
\caption{{\it Left:} In massive phase D5 brane wraps internal $S^2$ that shrinks to zero size in the interior at $z = z_m$: from lower dimensional point of view, brane simply ends. {\it Right:} Attached D3 brane wraps a smaller and smaller amount of the internal $S^3$ as it hangs deeper into the bulk. } 
\label{fig:braneclosed}
\end{figure}

\subsection{Monopole correlation}
Now from the discussion around Section \ref{sec:gapcharge}, we expect that if we have a charge gap the monopole operator should develop a vev, suggesting that the D3 brane should condense. It is interesting to see how this is realized geometrically. The key fact is that as the $S^2$ that the D5 wraps shrinks in the bulk, the D3 brane -- which ends on this $S^2$ -- also wraps a smaller and smaller portion of the $S^3$, as shown in Figure \ref{fig:braneclosed}. Its effective 4d mass is now position-dependent, decreasing in the bulk:
\begin{align}
R m(z) & = \frac{2 N_c}{\pi } \int_0^{\psi(z)} d\psi \sin^2 \psi \nonumber \\
& = \frac{ N_c}{\pi}\le[\cos^{-1}\le(\frac{z}{z_m}\ri)-\frac{z}{z_m} \sqrt{1 - \le(\frac{z}{z_m}\ri)^2}\ri] \ . 
\end{align}
Once the $S^2$ shrinks to zero size at $z_m$, the monopole worldline can simply end smoothly in the bulk at $z = z_m$. 

\begin{figure}[h]
\begin{center}
\includegraphics[scale=0.55]{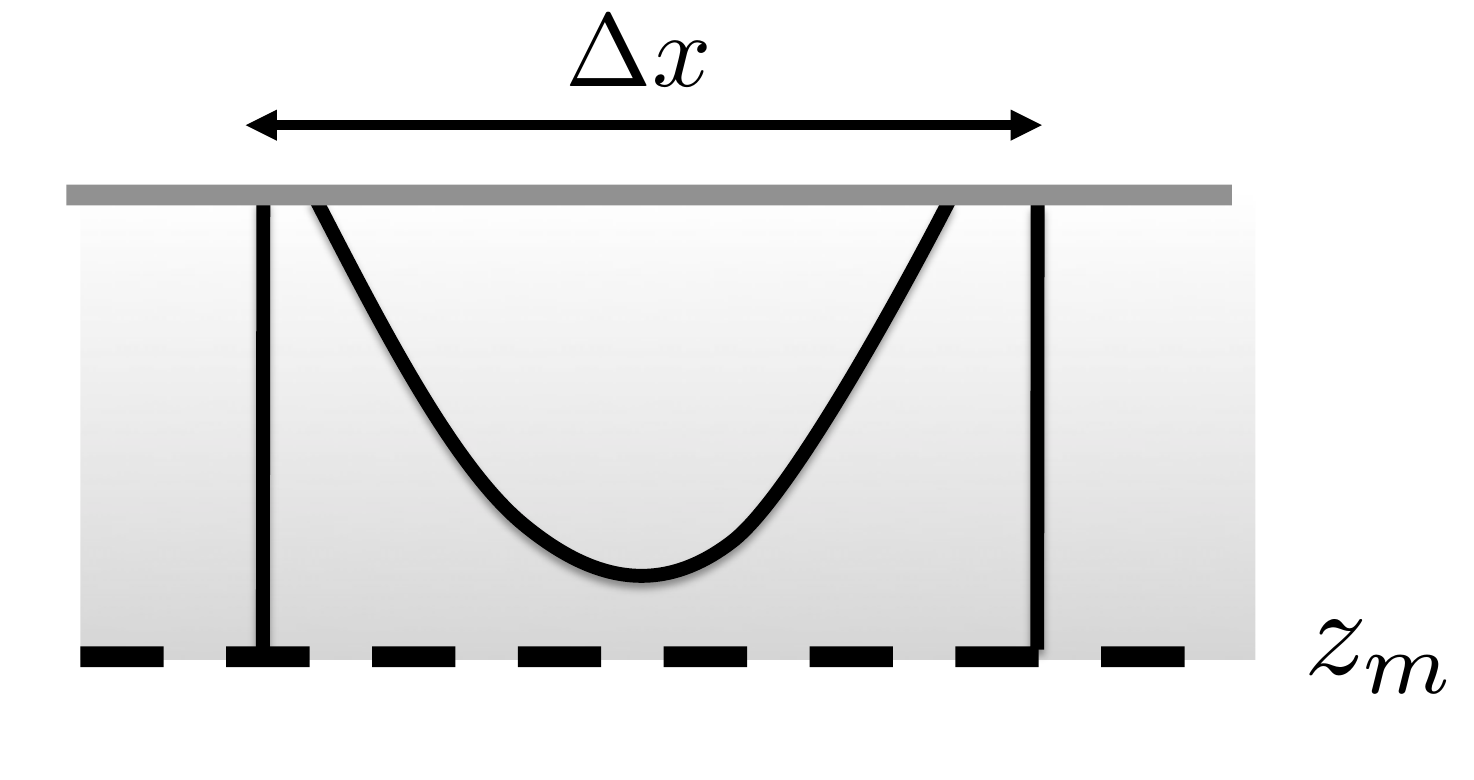}
\end{center}
\vskip -0.5cm
\caption{Possible configurations contributing to monopole correlator. For small $\Delta x$ the connected configuration dominates, whereas as $\Delta x$ is increased eventually bulk worldline breaks into two pieces. Monopole worldline is allowed to end at $z_m$ as it wraps a shrinking cycle.} 
\label{fig:disconnected}
\end{figure}

Thus the basic physics of the monopole two-point function can be seen without computation: at small separations the dominant configuration is a connected worldline whose action increases with distance, whereas at large separations the correlator will break into two disconnected pieces and the correlation function will saturate to a nonzero constant that is independent of distance, as shown in Figure \ref{fig:disconnected}. This is precisely the behavior expected in a phase where the monopole has condensed, as described in \eqref{monsat}. Each of the two disconnected pieces of the correlator should thus be understood as a vev for the monopole operator, and the fact that the worldline can {\it end} in the bulk should be understood as implementing the idea of a ``condensate'' -- normally thought of in terms of bulk fields -- in the worldline picture. We note that the presence of a wrapped shrinking cycle causing a topological transition is a familiar mechanism in holography, appearing in the holographic representation of the (Euclidean) screening of quarks at finite temperature \cite{Brandhuber:1998bs,Rey:1998bq} and the saturation of holographic entanglement entropy in a gapped phase \cite{Klebanov:2007ws}.

We now present some details of the (quite standard) computation. To obtain the effective action for the monopole, we start from the DBI action for the wrapped D3 brane and integrate over $(\psi, \th, \phi)$ to obtain
\be
S_{D3} =  \int_C ds\;m(z(s))\sqrt{G_{MN}(X(s))\dot{X}^M\dot{X}^N},
\ee
where $M,N$ run over the $AdS_4$ directions and $s$ parametrizes the remaining D3 brane direction along the worldline. An overdot denotes a derivative with respect to $s$. As $A = 0$ on this background the coupling \eqref{magcoup} plays no role in this analysis. We have assumed here that none of the transverse directions $X^M(s)$ depend on $\psi$, i.e. that the D3 brane is ``rigid'' in the way in which it wraps its half-$S^3$: we will not show that this is the most energetically favored configuration, and so this will remain an assumption in this note. We will discuss its consequences later. 

The simplest way to proceed is to define a conformally rescaled metric that takes into account the radial variation of the mass:
\be
\overline{G}_{MN}(x) \equiv G_{MN}(x) m(z)^2,
\ee
after which this takes the form of an ordinary geodesic action and we may use the usual machinery. 

We first solve for the connected solution: we take the geodesic to extend in the $x$ direction, and so we need to determine the curve $(x(s), z(s))$. There is a conserved momentum
\be
P \equiv \dot{x} \overline{G}_{xx},
\ee
and a constraint arising from reparametrization invariance with respect to $s$:
\be
\overline{G}_{xx} \dot{x}^2 + \overline{G}_{zz} \dot{z}^2 = 1
\ee
Using these relations, we solve for $\dot{z}$ and write the net change in $x$ and the total action $S$ in terms of the conserved momentum $P$:
\begin{align}
\Delta x & = 2 \int_\ep^{z_{\star}} dz \sqrt{\frac{\overline{G}_{zz}}{1 - \overline{G}^{xx} P^2}} \overline{G}^{xx} P \\
S_{D3} & = 2 \int_\ep^{z_{\star}} dz \sqrt{\frac{\overline{G}_{zz}}{1 - \overline{G}^{xx} P^2}},
\end{align}
where the turning point $z_{\star}$ is the solution to $\overline{G}^{xx}(z_{\star}) = \frac{1}{P^2}$, and $\ep$ is a UV cutoff. By varying $P$ and performing the integrals numerically we can find $S_{D3}[\Delta x]$. 

\begin{figure}[h]
\begin{center}
\includegraphics[scale=0.35]{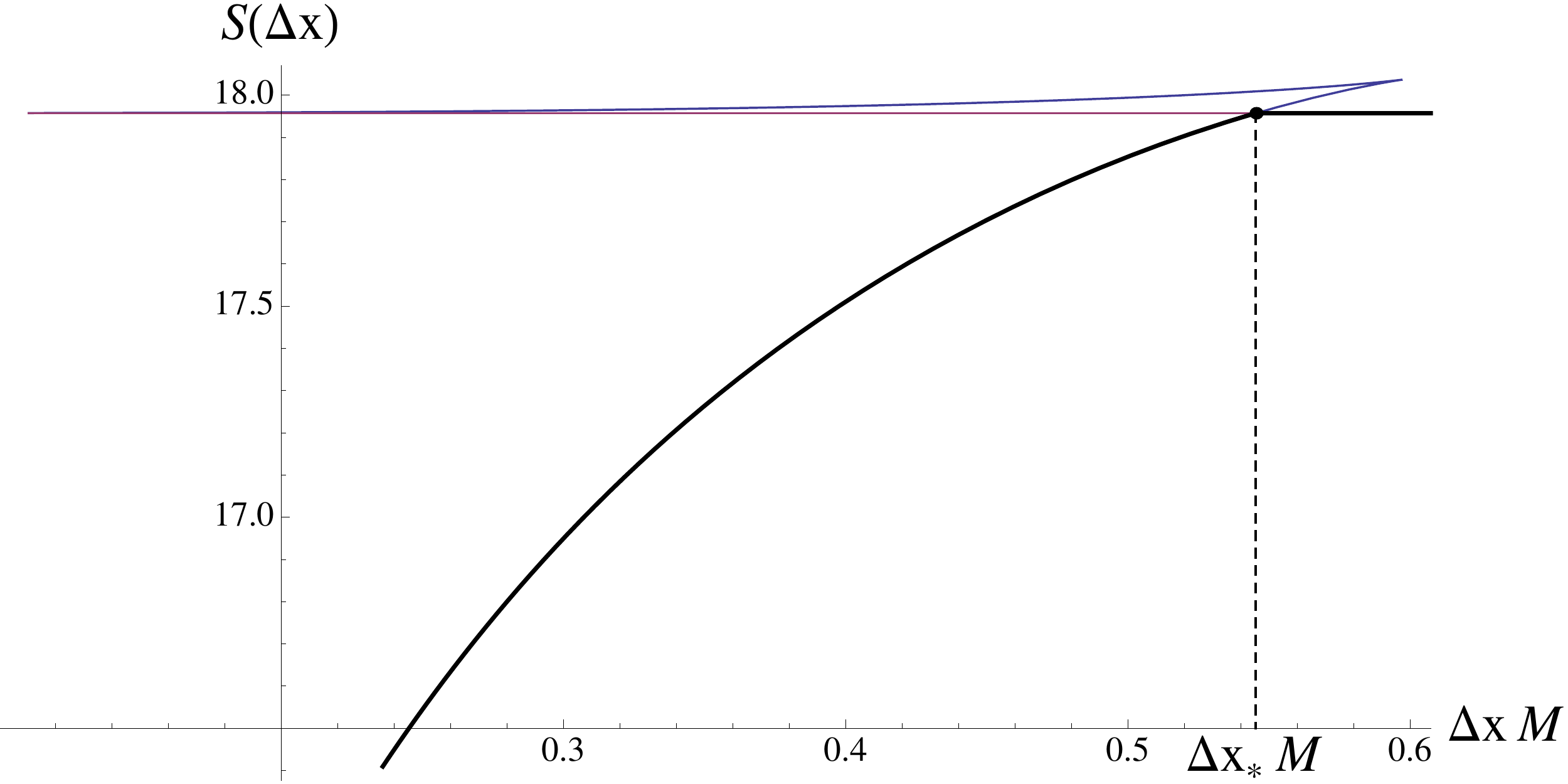}
\end{center}
\vskip -0.5cm
\caption{Monopole action on background with charge gap. Solid black line represents favored configuration; red and blue are unfavored branches of the disconnected (i.e. completely flat) and connected (i.e. swallowtail) configurations respectively. Note first-order transition at $\Delta x_{\star} M \approx 0.545$.}
\label{fig:massgeod}
\end{figure}

This action should be compared to that of the disconnected configuration, which is simply
\be
S_{disc} = 2\int_\ep^{z_m} dz \sqrt{\overline{G}_{zz}} 
\ee 
The results from such an analysis are shown in Figure \ref{fig:massgeod}, and display precisely the behavior predicted above. The critical value of $\Delta x$ of the phase transition is found numerically to be 
\be
\Delta x_{\star} M = 0.545\cdots \ .
\ee
If we relax the assumption that the transverse coordinates do not depend on $\psi$, then there may be an energetically preferred configuration that will dominate. This should affect physics on the scale of the AdS radius $R$, and thus may alter the precise value of $\Delta x_{\star}$, but we do not expect it to alter the qualitative shape of the curve shown in Figure \ref{fig:massgeod}, which is essentially measuring a geodesic length that probes scales much longer than the AdS radius. 

In this framework it is not possible to determine the precise normalization of the monopole condensate. However we may track its dependence on the mass: the full dependence on the mass comes from the UV logarithmic divergence in the integral, so we find
\be
\langle \sM \rangle \sim \exp\le(-\int_\ep^{z_m} dz \sqrt{\overline{G}_{zz}}\ri) \sim \le(M \ep\ri)^{\frac{N_c}{2}},
\ee
in agreement with considerations around \eqref{monsat}.

\subsection{Monopole condensation as confinement in the bulk}
We now pause to discuss the bulk interpretation of this calculation. On the worldvolume of the D5 brane lives a gauge field $A$: its dynamics is basically given by the Maxwell action, and so it is in a Coulomb phase. If the D5 brane caps off at $z = z_m$, from the 4d point of view, what happens to $A$ in the region $z > z_m$, where the D-brane simply does not exist? We are not actually allowed to simply delete a gauge field in a region of space: rather, we require an effectively 4d mechanism to remove it from the spectrum. 

\begin{figure}[h]
\begin{center}
\includegraphics[scale=0.55]{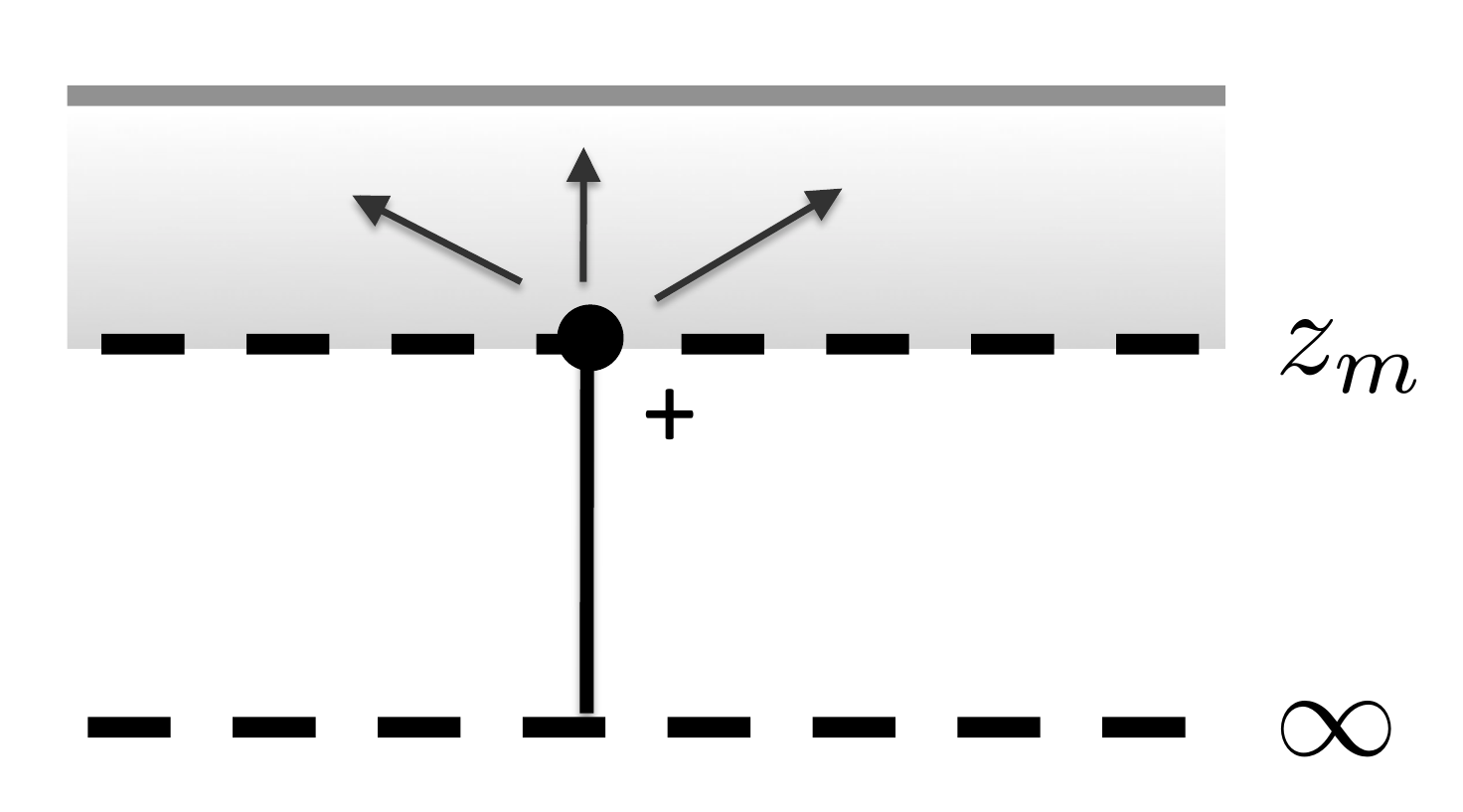}
\end{center}
\vskip -0.5cm
\caption{``Confinement'' in the bulk. For $z > z_m$ the D5 brane does not exist, and its worldvolume flux is confined into flux tubes (fundamental strings) which extend to the (Poincar\'e) horizon at $z = \infty$. From the 4d point of view it appears that the worldvolume gauge field has confined and monopoles have condensed.} 
\label{fig:confinement}
\end{figure}

In this case, the mechanism is {\it confinement}. For $z > z_m$ it is not true that the gauge field has ceased to exist; rather its electric flux is forced into tight flux tubes that we normally call fundamental strings. When these strings end on the D5 brane they cross into a deconfined phase and their flux is allowed to spill out into $A$, as shown in Figure \ref{fig:confinement}. The existence of such electric flux tubes is basically the definition of confinement. Indeed, confinement on D-brane worldvolumes and the subsequent realization of a fundamental string as an electric flux tube has been argued to play an important role in brane/anti-brane annihilation \cite{Yi:1999hd,Bergman:2000xf}, and the physics here is somewhat similar, except that the confinement is localized in space and (relatedly) should be thought of as confinement only from the effective 4d point of view. 

In four dimensions confinement is also associated with the condensation of magnetic monopoles, and indeed we have seen explicitly above that monopoles have condensed for $z > z_m$. Said slightly differently, there is no terribly good reason for the D5 brane to exist at all: it wraps an $S^2$ that is topologically trivial, and so the only thing stopping it from collapsing to a point is energetics, i.e. the fact that the slipping mode is dynamically stable. The magnetic monopole exploits this fact: looking at Figure \ref{fig:wrappedD3}, we see that it is a localized excitation that interpolates between an equatorial $S^2$ and a degenerate $S^2$ that has shrunk to a point. Thus the monopole costs some energy but uses it to collapse its parent D5 brane in its vicinity. A condensate of monopoles is equivalent to closing off the brane over a macroscopic region.

It is interesting that we were able to perform a controlled calculation to observe monopole condensation in the {\it bulk}, normally thought of as a strongly coupled phenomenon. Indeed it is true that the effective 4d gauge coupling
\be
\frac{1}{g_F^2} \sim \frac{N_c}{\sqrt{\lam}} \mbox{Vol}(S^2)
\ee
is blowing up as we approach $z = z_m$ from the D-brane side: however the full higher dimensional geometry is completely regular, a fact that we exploited in our calculation. It would be quite difficult to control such a calculation in a bottom-up context, and this is the main reason that we focused on a probe brane construction in this paper. This provides a simple holographic example of confinement in the bulk, and makes it particularly clear that {\it bulk confinement is dual to a charge gap on the boundary}, as articulated previously by various authors \cite{Faulkner:2012gt,Sachdev:2012tj} (see also \cite{Aharony:2012jf} for discussion of non-Abelian confinement in AdS). 

\section{Finite-density liquid} \label{sec:finiterho}
We now set the mass deformation to zero and turn instead to a finite density phase. Via the normal rules of AdS/CFT, the field theory charge density $\rho$ is equal to the boundary value of the bulk electric field, i.e. the canonical momentum $\rho(z)$ conjugate to the gauge potential $A_t(z)$:
\be
\rho(z) \equiv \frac{\delta S_{D5}}{\delta(\p_z A_t)} = \frac{2 N_c}{\pi \sqrt{\lam}} \frac{\p_z A_t}{\sqrt{1 - \frac{z^4 (2\pi)^2}{\lam} (\p_z A_t)^2}} \ . \label{rhodef}
\ee
As there is no charged matter in the bulk the equation of motion for $A_t$ is simply the bulk Gauss law:
\be
\p_z \rho(z) = 0
\ee
Thus we can evaluate $\rho(z)$ anywhere in the bulk, and it is always equal to the field theory charge density $\rho$, justifying our notational abuse. Note that a finite-density state requires that the brane always extend to the Poincar\'e horizon: there are no explicit bulk sources, and so the bulk electric field lines can only emerge from the horizon. This state and its generalizations with nonzero magnetic field and temperature have been extensively studied and are host to a rich body of physical phenomena, a sampling of which can be found in \cite{Jensen:2010ga,Evans:2010hi,Evans:2010iy,Myers:2008me,Filev:2009xp,Evans:2008nf}. 

Despite extensive study, this state remains somewhat exotic. It is a finite-density state that does not break the $U(1)_B$ symmetry (and so is not a superfluid) and yet at first glance also not appear to display the structure in momentum space required for a Fermi surface (and so is not a Fermi liquid). Connecting such holographic phases to a more conventional understanding of states of quantum matter remains an important and open question. 

It is nevertheless of interest to understand how the monopole correlation behaves on this background. The finite charge density has an important effect that is easy to understand: the magnetic monopole wants to move in dual ``Landau levels'' of the background electric field. Landau level wavefunctions are Gaussian in space, and thus we expect the correlation to fall off in a similar fashion. This is precisely what happens, and in the remainder of this section we will derive this result through a Euclidean world-line calculation. 

\subsection{Conserved quantities on magnetic backgrounds}
The effective Euclidean action for the monopole takes the form
\be
S_{D3} = m \int ds + i q_m \int \tA, \label{effacB}
\ee
The first term is a proper length on $AdS_4$ and the second is a coupling to $\tA$, the magnetic dual of $A$, as shown in \eqref{magdual}. We will keep $m$ and $q_m$ arbitrary in this section as the analysis does not require any stringy ingredients. To understand the factor of $i$ we note that on a real trajectory the coupling to a background gauge field always generates phases, even in Euclidean signature. Now evaluating \eqref{magdual} explicitly and using the fact that the only nonzero component of $F$ is $F_{zt} \sim \rho$, we see from \eqref{rhodef} that $d\tA$ takes a very simple form:
\be
d\tA = \rho\;dx \wedge dy \ . 
\ee
We need to pick a gauge for $\tA$: for concreteness we will use 
\be
\tA = -\rho y\;dx, \label{tAgauge}
\ee
and will comment on the (easily understandable) gauge-dependence of our answers at the end. It is clear that the problem we are solving is equivalent after a change of notation to that of an {\it electric} charge moving in a background {\it magnetic} field. 

We now need to solve the geodesic equation following from \eqref{effacB}:
\be
m \frac{D^2 X^{M}}{ds^2} - iq_m (d\tA)^{M}_{\phantom{M}N} \dot{X}^N = 0 \label{geodF}
\ee
This is generally done by finding conserved quantities associated to Killing vectors $\xi^{\mu}$ of the background metric, e.g. translation along the spatial directions. The situation is slightly more complicated here: from \eqref{tAgauge} we see that $\tA$ does not appear invariant under spatial translations. It is, however, invariant up to a $U(1)$ gauge transformation, and one would hope that this would be just as useful. 

Generalizing slightly, denote by $\sL_{\xi}$ the Lie derivative along $\xi$ and suppose we have a background $(G_{MN}, \tA_M)$ such that
\be
\sL_{\xi}(G_{MN} ) = 0 \qquad \sL_{\xi}(\tA_M ) = \p_M \Lam_{\xi}
\ee 
with $\Lam_{\xi}$ a scalar function of spacetime. Due to the gauge-invariance of the action \eqref{effacB} this is a symmetry: thus by applying the usual Noether procedure to \eqref{effacB} with the infinitesimal transformation $\delta X^{\mu} = \xi^{\mu}$ we find the following conserved quantity along the geodesic:
\be
Q_{\xi}(s) \equiv \xi_{M} \le(m \dot{X}^M(s) + i q_m \tA^M(X(s))\ri) - i q_m \Lam_{\xi}(X(s)) \ . \label{Qdef}
\ee
We can also verify directly from \eqref{geodF} that $\frac{d}{ds}Q_{\xi} = 0$ along the geodesic. We see that there is still a conserved charge, but it explicitly involves the gauge transformation parameter $\Lam_{\xi}$. 

\subsection{Geodesic action}
With these conserved quantities it is easy to construct the on-shell action. We seek a geodesic that intersects the boundary at $(x,y) = \le(\frac{\pm \Delta x}{2}, 0\ri)$. The geodesic will move in $(x,y,z)$. Using \eqref{Qdef} we can construct conserved quantities associated with spatial translations:
\begin{align}
P_{x} & = m \dot{x} G_{xx}(z) - i q_m \rho y \nonumber \\ 
P_{y} & = m\dot{y} G_{xx}(z) + i q_m \rho x \label{consmot}
\end{align}
In the gauge \eqref{tAgauge} translational invariance in the $x$ direction is manifest, but that in the $y$ direction requires the usage of the formalism developed above. The final expressions are pleasingly symmetric. 

We will also use the normalization of the four-velocity
\be
G_{xx}\le(\dot{x}^2 + \dot{y}^2\ri) + G_{zz} \dot{z}^2 = 1 \ . \label{fournorm}
\ee
It is convenient to define a new parameter $\lam$ along the path
\be
G_{xx}(z(s))\frac{d}{ds} = \frac{d}{d\lam} \label{lamdef}
\ee
With the constants of motion \eqref{consmot} we can immediately solve for the dependence of $x$ and $y$ on $\lam$:
\begin{align}
x & = x_0 + \al \cosh(\om \lam) + \beta \sinh(\om \lam) \nonumber \\
y & = y_0 - i\le(\al \sinh(\om \lam) + \beta \cosh(\om \lam)\ri) \mbox{sgn}(q_m \rho) \label{xysol}
\end{align}
where $\om$ is the cyclotron frequency, defined so that it is always positive
\be
\om = \frac{|q_m \rho|}{m} \label{cyc}
\ee
and $x_0, y_0, \al, \beta$ are free integration constants. Note that if the displacement in $x$ is real, then that in $y$ is necessarily imaginary: the geodesic that we are constructing will be complex. 

Take $\lam = 0$ to be the midpoint of the geodesic: then our boundary conditions require $x_0 = \alpha = 0$. It remains to relate $\beta$ and $y_0$ to the boundary conditions. Now plugging into \eqref{fournorm} we find
\be
G^{xx}(z) (\om\beta)^2 + G_{zz}(z) \dot{z}^2 = 1
\ee
Using the $AdS_4$ metric \eqref{AdS4met} we can solve for $\frac{dz}{ds}$,
\be
\frac{dz}{ds} = \frac{z}{R} \sqrt{1 - \le(\frac{z}{z_*}\ri)^2} \qquad z_* \equiv \frac{R}{\om\beta} \ . 
\ee
Now we can use \eqref{lamdef} to determine the change in $\lam$ along the path; if we have $\lam = 0$ at the turning point, then at the initial endpoint we find
\be
\lam_i = -\int_0^{z_*} dz \frac{ds}{dz} G^{xx} = -\frac{R}{(\om\beta)^2}
\ee
Putting this into \eqref{xysol} we find the desired relation between $\beta$ and $\Delta x$:
\be
\beta\sinh\le(\frac{R}{\om\beta^2}\ri) = \frac{\Delta x}{2} \label{betaexp}
\ee
This cannot be explicitly solved for $\beta$; however in the large $\Delta x$ limit $\beta$ is reasonably well-approximated by
\be
\beta(\Delta x \to \infty) \approx \zeta\sqrt{\frac{m R}{\log\le(\frac{\Delta x}{\zeta\sqrt{m R}}\ri)}}\ . 
\ee
where we have used \eqref{cyc} and defined a correlation length $\zeta$ as
\be
\zeta = \frac{1}{\sqrt{|q_m \rho|}}
\ee
What does $\zeta$ measure? If the Dirac condition is saturated we have $q_m = \frac{2\pi}{q_e}$ and the factor $(q_m \rho)^{-\ha}$ is then a length scale characterizing the area occupied by a single field-theoretical quantum of charge in the finite-density state. In a Fermi liquid this scale would correspond to the inverse Fermi momentum, $\zeta = \sqrt{2}k_F^{-1}$. Such a length scale can emerge from a holographic calculation only because of the nonperturbative nature of the magnetic monopole, whose magnetic charge knows about the fundamental electric charge quantum $q_e$ \cite{Faulkner:2012gt}. 

We still need to fix $y_0$: as we would like the geodesic to start and end at $y = 0$, we pick $y_0 = i\beta \cosh(\om \lam_i)$. This completely fixes the solution. 

We now compute the on-shell action \eqref{effacB} in the large $\Delta x$ limit. The geometric portion (at large $\Delta x$) is
\be
m \int ds \approx 2 m R \log\le(\frac{\zeta}{\ep}\sqrt{m R \log\le(\frac{\Delta x}{\zeta\sqrt{m R}}\ri)}\ri), \label{geom}
\ee
with $\ep$ a UV cutoff. Note that it depends very weakly on $\Delta x$: essentially the geometric portion saturates at the length scale given by $\zeta$.

The contribution from the coupling to $\tA$ is more interesting. Putting in the explicit solutions \eqref{xysol} and using the form \eqref{tAgauge} we find at large $\Delta x$ 
\be
i q_m \int dx \tA_x \approx \frac{|q_m \rho|}{4} \beta^2 \exp\le(\frac{2}{\om \beta^2}\ri),
\ee
which can be further simplified using \eqref{betaexp} to be
\be
i q_m \int \tA \approx \frac{|q_m \rho|}{4}(\Delta x)^2  = \frac{1}{4}\le(\frac{\Delta x}{\zeta}\ri)^2. 
\ee
It is interesting to note that the seemingly imaginary coupling has resulted in a {\it real} contribution to the action. This is because the geodesic was also complex, moving in the imaginary $y$ direction, meaning that we pick up an extra factor of $i$ from the evaluation of the gauge field \eqref{tAgauge}. The result may be thought of as a WKB derivation of the familiar Landau level wavefunction (if in curved space). 

We may now assemble the pieces. At large $\Delta x$ the geometric part \eqref{geom} depends very weakly on spatial separation and can be neglected. Thus we conclude that the correlator behaves as
\be
\langle \sM(\Delta x) \sM^{\dagger}(0) \rangle \sim \exp\le(-\frac{|q_m\rho|}{4}(\Delta x)^2\ri) \label{gausscorr}
\ee
In the specific case of the D3 brane one should set $q_m = 2\pi$ as in \eqref{magcoup}. This computation of the monopole correlation in a compressible finite-density phase is one of the main results of this paper. We see that the correlation is very strongly suppressed in space.

Finally, we should discuss the monopole correlation in (Euclidean) time. If we separate the endpoints of the geodesic in time, it does not couple to the background electric field, and the calculation is the same as in pure $AdS_4$, i.e. 
\be
\langle \sM(\Delta t) \sM^{\dagger}(0) \rangle \sim (\Delta t)^{-2\Delta},
\ee
with $\Delta = mR$. Thus the correlation is power-law in time and Gaussian in space. 

We briefly comment on the gauge-dependence of our answer. Any charged correlator in a background field depends on the gauge. Gauge transformations generated by a gauge parameter $\Lam(x)$ will shift the answer by \eqref{gausscorr} by a phase factor $\exp(iq_m(\Lam(\Delta x) - \Lam(0)))$. Similarly, if we translate or rotate the correlator \eqref{gausscorr} we will generally pick up phases, as the gauge choice \eqref{tAgauge} will no longer exactly line up with the interval choice of a separation purely in the $x$-direction. The absolute value of the correlator will not change. 

\section{Conclusion} \label{sec:conc}
In this work we computed monopole correlation functions from holography by relating a boundary monopole operator to a bulk magnetically charged excitation. We worked largely in the context of a particular top-down brane model, the D3D5 intersection, where the bulk monopole is a type of wrapped D3 brane. We demonstrated that in a phase when the boundary $U(1)$ current $j^{\mu}$ is gapped, the bulk monopole can be thought of as being {\it condensed} in the region of space where the D5 brane does not exist. From a 4d point of view, this looks like bulk confinement of the worldvolume gauge field, with the fundamental string being the flux tube. Despite its somewhat exotic bulk realization, this condensation is precisely what is expected from a gapped phase on field-theoretical grounds. 

We also presented results on monopole correlations in a finite-density compressible phase. We found the correlation to die off as a Gaussian in space in \eqref{gausscorr} -- this follows simply from an understanding of the bulk magnetic monopole moving in dual ``Landau levels'' of the bulk electric field. 

Some aspects of the calculation presented here do not really require holography and can be understood in any system with a $U(1)$ current $j^{\mu}$. As it is conserved, we may write it as the curl of an auxiliary gauge field 
\be
j^{\mu} \equiv \frac{i}{4\pi} \ep^{\mu\nu\rho}\p_{\nu} b_{\rho} \label{jbrels}
\ee
Now the coupling between the source gauge field $a_{\mu}$ and $j_{\mu}$ is (after an integration by parts):
\be
\int d^3x\;a_{\mu}j^{\mu} = \frac{i}{4\pi}\int d^3 x\;\ep^{\mu\nu\rho}(\p_{\mu} a_{\nu}) b_{\rho} = \int d^3x\;b_{\mu} \topj^{\mu} \label{topcoups}
\ee
In words, in any system the topological current $\topj$ under which the monopole is charged is coupled to an external gauge field source $b$ which is related to the {\it ordinary} current via \eqref{jbrels}. If $j$ has a nonzero charge density, then $b$ corresponds to an applied magnetic field, and presumably the monopoles which make up the current ``feel'' this applied field. This is clearly the physics that is encoded holographically in the coupling \eqref{magcoup}. 

We see that the key ingredient from holography was not the coupling between the charge density and the monopole, but rather the relation of the monopole to a gapped excitation in the bulk whose dynamics could be easily understood. It is this identification which fails in a conventional Fermi liquid, where a monopole operator does not obviously create a well-defined object (see however \cite{2011PhRvB..84p5126M} for a discussion of vortex-like objects in Fermi liquids). It would be  very interesting to compute the monopole correlation in a Fermi liquid, perhaps through an extension of the techniques in \cite{PhysRevB.78.085129}. We note that in a superfluid with a net charge density $\rho$ the monopole creates a gapped vortex (now in $(2+1)$ dimensions). If the vortex is heavy we can compute its two-point function in a calculation similar to that above, except in (2+1) dimensional flat space, and we find precisely the same Gaussian suppression.   

We briefly comment on the relation between the calculations in this paper and the RPA approach to monopole correlations described in Section \ref{sec:FL}. Such an approach takes into account only the energy stored in the gauge field: if applied to our system it would entirely miss the tension of the monopole worldline, a tension that encodes the fact that the monopole is a nonperturbative object. In fact if we expand the DBI action about the conformal phase and examine the normalization of the Maxwell kinetic term we find:
\be
S = \frac{N_c}{4 \pi \sqrt{\lam}}\int d^4 x\;(dA)^2,
\ee
The normalization of the Maxwell action maps to that of the current-current correlator\footnote{This statement requires a choice of normalization for the boundary theory current. In this section we discuss the normalization provided by string theory, but actually similar statements can also be made in a bottom-up model, where the key physics is just that the bulk monopole should be sufficiently heavy that quantum corrections to its mass can be neglected. This is the case in models that remain weakly coupled in the bulk. } and so if we were to perform an RPA-type analysis as in \eqref{effA} we would find a contribution with an overall scaling $\sO(\frac{N_c}{\sqrt{\lam}})$. This should be compared to the $\sO(N_c)$ dependence of the monopole worldline mass, and so is subleading at strong coupling. The presence of the extra parameter $\lam$ in this theory lets us cleanly separate these two contributions.

We now discuss some directions for future research. The monopole correlation function can easily be studied holographically in more general circumstances, e.g. at finite temperatures or near phase transitions \cite{Mateos:2006nu,Kobayashi:2006sb,Albash:2006ew,Mateos:2007vn}, such as that between Minkowski and black hole embeddings at finite baryon chemical potential and quark mass. We note also that the monopoles studied here -- which try to close off the bulk D-brane -- seem to be the conjugate objects to the worldsheet instantons studied in \cite{Faulkner:2008qk}, which try instead to pull the bulk D-brane into the black hole horizon. Such objects were argued to play an important role in the aforementioned phase transition at finite $\lam$ \cite{Faulkner:2008hm}, and it would be very interesting to understand if (and how) the monopoles studied here interact with that story. 

Many of the considerations studied in this note may be extended to higher dimension, and again holography may be helpful for simple computations. For example, in a general $d$-dimensional field theory, the analog of a monopole operator is a $d-3$-dimensional object $\Sigma$ along which $f$ is not closed:
\be
d f = \delta^{(3)}(``x-\Sig")
\ee
To take the familiar case of $(3+1)$ dimensions, the monopole operator becomes an extended one-dimensional object, a t'Hooft line. In a phase with gapped charges, a closed t'Hooft loop will obey a perimeter law, which is the analog of the factorization of the monopole correlator in a gapped phase in \eqref{monsat}\footnote{The notion of ``generalized global symmetries'' recently introduced in \cite{Kapustin:2014gua} is useful here and is the extension of the topological symmetry $\topj$ to higher dimension.}.

Finally, we close on a more speculative note. Holographic phases of compressible matter are somewhat mysterious: they do not fit easily into a textbook classification of quantum matter, essentially because they can support a finite density of charge without displaying the structure in momentum space that is normally associated with a Fermi surface \cite{Hartnoll:2011fn,Huijse:2011hp,Sachdev:2010uz,Iqbal:2011in}. The situation is slightly less mysterious in a $(2+1)$ dimensional bulk. It was shown in \cite{Faulkner:2012gt} that essentially any holographic compressible state in (2+1) dimensions will exhibit {\it Friedel oscillations} in its density correlation function. Such oscillations are normally associated with the presence of a Fermi surface, but in the holographic model it arose from the presence of a dilute gas of {\it bulk} magnetic monopole instantons. This calculation has not yet been extended to higher dimensions, but intuition arising from the low-dimensional calculation suggests that magnetically charged objects are likely to be important in any sufficiently fine-grained description of holographic matter, even in higher dimension \cite{Sachdev:2012tj}. While further study is required, we hope that the monopoles studied here will eventually help build a bridge between a conventional description of compressible phases and that provided by holography.  

\vspace{0.2in}   \centerline{\bf{Acknowledgements}} \vspace{0.2in} I thank M.~Barkeshli, D.~Das, T.~Faulkner, T.~Grover, S.~Hartnoll, D-K.~Hong, K-S.~Kim, M.~Metlitski, M.~Mezei, S.~Sachdev, J.~Polchinski, A.~Puhm, D.~Tong, T.~Ugajin, and P.~Yi for helpful discussions and correspondence. I also thank the organizers of the Amsterdam String Workshop 2014 and the APCTP Focus Program on Aspects of Holography for their hospitality and for giving me the opportunity to present this work in stimulating environments. I was supported in part by the NSF under Grant No. PHY11-25915 and by the DOE under Grant No. DE-FG02-91ER40618. 

\begin{appendix}

\section{Derivation of monopole couplings} \label{app:effac}
The bulk monopole studied in this paper is actually a D3 brane that wraps a hemispherical $S^3 \subset S^5$ and ends on the D5 brane in a manner described in detail in Section \ref{sec:monop}. In this Appendix we derive its couplings to the worldvolume fields living on the D5 brane. 

The starting point is the following action:
\begin{align} 
S =  & -T_5 \int_{D5} d^6 \sig \sqrt{-\det(\ga + 2\pi \al' F)} \nonumber \\ & + T_5 \int_{D5} 2\pi \al' F \wedge C_4 + T_3 \int_{D3} C_4 + \cdots
\end{align}
$F$ is the field strength of the D5 brane worldvolume gauge field, $F = dA$. In this action the electric gauge field $A$ is the dynamical variable, but we know that the edge of the D3 brane should couple magnetically to $A$, and so we need to dualize this action. This is done by treating $F$ as the dynamical variable rather than $A$ and supplementing the action with a 3-form Lagrange multiplier $K_3$ to guarantee that $dF = 0$ everywhere except on the edge of the D3 brane:
\be
S_K = \int_{D5} K_3 \wedge dF + q_m \int_{\p D3} K_3 \label{lagac}
\ee
Now the portion of the action involving couplings to the form fields is
\begin{align}
S + S_K & =  \int_{D5} F \wedge \le(d K_3 + 2\pi \al' T_5 C_4\ri) \nonumber \\ & + T_3 \int_{D3} C_4 + q_m \int_{\p D3} K_3 + \cdots \label{CScoups}
\end{align} 
If we proceed to eliminate $F$ from this action using its equation of motion then $K_3$ will be its 6d magnetic dual. Now $C_4$ has a gauge transformation under a 3-form gauge parameter $\Lam_3$. For the first term above to be gauge-invariant we see that we require that $K_3$ also transform under $\Lam_3$:
\be
\delta_{\Lam} C_4 = d\Lam_3 \qquad \delta_{\Lam} K_3 = -2\pi \al' T_5 \Lam_3 \ . \label{c4trans}
\ee 
The sum of the second two terms in \eqref{CScoups} will also be gauge invariant if we pick $q_m$ to be
\be
q_m = \frac{T_3}{2\pi \al' T_5} = 2\pi \ . 
\ee
The fact that this coupling is nonzero means that the edge of the D3 brane couples to $K_3$. Now to interpret this from the 4d point of view we take the following ansatz for $K_3$:
\be
K_3 = \frac{1}{4\pi} \tA \wedge V_2
\ee
with $V_2$ the volume form of a unit $S^2 \subset S^5$. The edge of the D3 brane is a product of this $S^2$ and a one-dimensional curve $C \subset AdS_4$, and  integrating over the $S^2$ we find from \eqref{lagac}:
\be
S_K = \int_{AdS_4} F \wedge d\tA + q_m \int_{C} \tA
\ee 
$\tA$ is the 4d magnetic dual to $A$, and the second term is the desired coupling between the monopole (effectively a particle moving along the worldline $C$ on $AdS_4$) and $\tA$. To understand the precise relation between $\tA$ and $A$ we vary the total action with respect to $F$ to find:
\begin{align}
& (d \widetilde{A})_{MN} \nonumber \\
& = \sN \sqrt{-\det_4\le(\ga + 2\pi \al' F\ri)}[(\ga + 2 \pi \al' F)^{-1}]^{PQ}\ep_{PQMN} 
\end{align}
where the normalization $\sN = -4\pi^2 \al' R^2 T_5$, as claimed in \eqref{magdual}. This is the generalization of the idea of a magnetic dual to the nontrivial gauge kinetic term in the DBI action. At small $F$ it reduces to the more familiar $d\tA \sim \star_4 F$ of Maxwell electric-magnetic duality. 
\section{Transformation of monopole under global symmetries} \label{app:global}
Recall that the global symmetry group of the defect field theory is $U(1)_B \times SU(2)_V \times SU(2)_H$, where the two $SU(2)$ factors are realized geometrically as the isometry groups of the two $S^2$ factors inside the $S^5$. The fermions are charged in the fundamental under $SU(2)_V$ (in which the D5 brane is not extended), and the scalars are in the fundamental under $SU(2)_H$ (which the D5 brane wraps). In this section we seek to understand how the monopole operator transforms under this global symmetry group. 

We start from the field theory, at zero coupling. We essentially follow the arguments given in \cite{Borokhov:2002ib}, with slight modifications to deal with the different symmetry group in our problem. To determine the quantum numbers of the monopole at large $N_c$ we can study the field theory on $S^2 \times \mathbb{R}$ with a classical gauge field background corresponding to a single unit of $U(1)_B$ flux on the $S^2$. As we are interested in the operator with lowest conformal dimension, we would like to understand the ground state of this system. 

However there is some ambiguity as to what we mean by ``ground state'', due to the existence of fermion zero modes bound to the monopole. There are $2 N_c$ zero modes, one for each fermion species, and we denote them by $c^a_{i}$, where $a$ is a $SU(N_c)$ color index and $i$ an $SU(2)_V$ spinor index. The only (color) gauge-invariant combination of zero mode creation operators takes the baryonic form
\be
C_{i_1 i_2 \cdots i_{N_c}} = \ep_{a_1 a_2 \cdots a_{N_c}} (c^{a_1}_{i_1})^{\dagger} (c^{a_2}_{i_2})^{\dagger} \cdots (c^{a_{N_c}}_{i_{N_c}})^{\dagger},
\ee
with $\ep$ representing the antisymmetric symbol. This object is entirely symmetric in its spin indices and so transforms in the $N_c+1$ dimensional spin-$\frac{N_c}{2}$ representation under $SU(2)_V$. 

Denote the Fock vacuum -- which is annihilated by all the $c$'s -- by $|0\rangle$. To create a gauge-invariant state we must act with the $C$'s defined above. Acting with them once we find the $N_c+1$ states described above:
\be
C_{i_1 i_2 \cdots i_{N_c}} |0\rangle \label{zerocharge} \ . 
\ee
Acting with them twice the only non-vanishing state is
\be
C^2|0 \rangle \equiv \ep^{i_1 j_1}  \cdots \ep^{i_{N_c} j_{N_c}} C_{i_1\cdots i_{N_c}} C_{j_1 \cdots j_{N_c}}|0\rangle \ . 
\ee
This state has every zero mode filled and is a singlet under $SU(2)_V$, just like $|0\rangle$. Each $C$ carries $U(1)_{B}$ charge $N_c$: this fixes the relative charge assignments, but there is an ambiguity in deciding which state should carry $U(1)_B$ charge $0$. It is argued in \cite{Borokhov:2002ib} that a CP-invariant quantization requires that we assign equal and opposite $U(1)_B$ charge to $|0\rangle$ and $C^2|0\rangle$. This prescription gives $|0\rangle$ and $C^2 |0 \rangle$ a $U(1)_B$ charge of $-N_c$ and $N_c$ respectively, and states that the states \eqref{zerocharge} carry zero $U(1)_B$ charge. Presumably these are the states we have been studying in this note.

This analysis was all performed at zero coupling. As we go to strong coupling, it is not clear how much of the physics of these zero modes should survive: in particular, it seems that interactions could modify the energies of the charged monopole states $|0\rangle$ and $C^2|0\rangle$ relative to the charge $0$ states \eqref{zerocharge}. However as the states in \eqref{zerocharge} form an irreducible multiplet under $SU(2)_V$ they should remain degenerate, and so even at strong coupling we expect the charge-$0$ monopole to transform as a spin-$\frac{N_c}{2}$ representation under $SU(2)_V$. 

Interestingly, it is actually possible to reproduce this result from a holographic analysis, to which we now turn. Consider the metric of a unit $S^5$: 
\begin{align}
ds_{S^5}^2 = d\psi^2 & + \sin^2\psi\le(d\th^2 + \sin^2 \th d\phi^2\ri) \nonumber \\ 
& + \cos^2 \psi\le(d\tilde{\th}^2 + \sin^2 \tilde{\th} d\tilde{\phi}^2\ri) \ 
\end{align}
The monopole D3 brane wraps $(\psi, \th, \phi)$ and sits at a point at $(\tth, \tphi)$. $SU(2)_V$ acts nonlinearly as a rotation on $(\tth, \tphi)$, and so it appears that that the existence of the D3 brane at a definite location spontaneously breaks $SU(2)_V$ entirely. This cannot be true at the quantum mechanical level. In the finite-$N_c$ field theory the D3 brane is associated with a pointlike operator, and the Coleman-Mermin-Wagner-Hohenberg theorem states that the breaking of a continuous symmetry can happen only on an object with at least two extended dimensions \cite{Coleman:1973ci,PhysRevLett.17.1133,PhysRev.158.383}. In other words, we should not be able to assign $(\tth,\tphi)$ a classical value: rather the wavefunction of the D3 brane in the bulk is delocalized in the $(\tth, \tphi)$ directions.  

To understand this wavefunction, we should quantize these two directions. Following the discussion in the field theory, we study the boundary field theory on $S^2 \times \mathbb{R}$ with a unit of flux and attempt to understand the Hilbert space of the monopole. In the bulk we have global $AdS_4$ with the monopole at rest at the origin in the interior of the geometry. The only low-lying fluctuations are in $(\tth, \tphi)$: we study a configuration where these fluctuations depend only on time and thus trace out a worldline on the $S^2 \subset S^5$ representing $SU(2)_V$. The relevant action is
\be
S = -T_3 \int d^4\sig \sqrt{-\det\ga_3} + T_3 \int_{D3} C_4 \ . 
\ee
The eigenvalues of the Hamiltonian governing this worldline quantum mechanics are related to the dimensions of a tower of operators associated with the monopole using the usual rules of AdS/CFT. We are interested in the lowest-energy state. 

The coupling of the monopole D3 brane to the background $C_4$ sourced by the color D3 branes plays an interesting role here. The portion of $C_4$ with legs along the $S^5$ can be taken to be:
\be
C_4 = 8\lam l_s^4 \sin^2 \psi \cos^2 \psi d\psi \wedge d(\cos \th) \wedge d\phi \wedge b + \cdots
\ee
where $b$ is a one-form chosen so that $db$ is proportional to the volume form on $S^2$:
\be
db = \ha \sin \tth d\tth \wedge d\tphi \ . 
\ee
This means that $b$ can be interpreted as the gauge potential for a unit-charge magnetic monopole\footnote{We stress that there are multiple kinds of monopoles involved here. There is the monopole in $U(1)_B$ that we have been studying for this entire paper, which creates flux in the field-theory directions. This is distinct from the monopole in the background field $b$, which is a component of the $C_4$ sourced by the color D3 branes, and creates flux in the compact string-theory directions.} at the core of this $S^2$. Then letting the index $\al$ run over $(\tth, \tphi)$ and integrating over all compact directions we find that the coupling to $C_4$ becomes,
\be
T_3 \int_{D3} C_4 = N_c \int b_{\al} \frac{dX^{\al}}{dt} dt \ .  \label{moncoup}
\ee
This looks like the coupling of a charged particle to a background $U(1)$ gauge field. Now expanding out the DBI action in derivatives we find
\begin{align} 
- T_3 \int & d^4\sig \sqrt{-\det\ga_3} = \nonumber \\
-& \frac{N_c}{2 R} \int dt \le( 1 - \frac{R^2}{8} \le(\dot{\tth}^2 + \cos^2 \tth \dot{\tphi}^2 \ri)\ri) + \cdots \label{kinterms}
\end{align} 
The first term is a classical energy corresponding to the rest mass of the particle, as in \eqref{D3mass}. The second term gives dynamics to the $S^2$ coordinates. 

Combining \eqref{moncoup} and \eqref{kinterms} we find that the system is given by a charged particle moving on an $S^2$ with a magnetic monopole of charge $N_c$ at its core. This is a well-studied action, and is in fact the path-integral representation of a spin-$\frac{N_c}{2}$ system (for a pedagogical review see e.g. Chapter 7 of \cite{fradkinftcmp}). To be more precise, the lowest Landau level of the particle on the $S^2$ has degeneracy $N_c +1$ and transforms as a spin-$\frac{N_c}{2}$ representation under $SU(2)_V$. Pleasingly, this is precisely the representation expected for the monopole operator on field-theoretical grounds. It is interesting to see how the fermion zero modes are geometrized into the quantum mechanics of the collective coordinate on $S^2$. 

From the prefactor of the kinetic term in \eqref{kinterms} we see that the spacing between this degenerate ground state and the other bulk energy levels scales like $(N_c R)^{-1}$. As we take $N_c \to \infty$ all of these states become degenerate, collapsing onto the ground state and permitting spontaneous symmetry breaking of $SU(2)_V$. There is some similarity between this discussion and that of giant gravitons \cite{McGreevy:2000cw}, which is different in detail but where the same coupling between probe branes and background flux also allows access to physics that is nonperturbative in $N_c$.

\end{appendix}
\bibliographystyle{utphys}
\bibliography{all}

\end{document}